\documentclass[fleqn,usenatbib]{mnras}
\usepackage{fix-cm}
\usepackage{xcolor}

\usepackage{newtxtext,newtxmath}
\usepackage{booktabs,graphicx}
\usepackage{multirow}

\usepackage[T1]{fontenc}

\DeclareRobustCommand{\VAN}[3]{#2}
\let\VANthebibliography\thebibliography
\def\thebibliography{\DeclareRobustCommand{\VAN}[3]{##3}\VANthebibliography}

\usepackage{graphicx}	
\usepackage{amsmath}	

\title[Star-forming Fraction in BCGs]{Galaxy Clusters from the DESI Legacy Imaging Surveys. 
 III. Star-forming Fraction of Brightest Cluster Galaxies }

\author[S. Liu et al.]{Shufei Liu$^{1,2}$,
Hu Zou$^{1,2}$\thanks{E-mail:zouhu@nao.cas.cn}, 
Jinfu Gou$^{1,2}$, 
Weijian Guo$^{1}$, 
Niu Li$^{1}$, 
Wenxiong Li$^{1}$, 
Gaurav Singh$^{1}$, 
\newauthor
Haoming Song$^{1,2}$, 
Jipeng Sui$^{1,2}$, 
Xi Tan$^{1,2}$, 
Yunao Xiao$^{1,2}$, 
Jingyi Zhang$^{1}$,
Lu Feng$^{1}$
\\
$^{1}$National Astronomical Observatories, Chinese Academy of Sciences, Beijing 100101, China\\
$^{2}$School of Astronomy and Space Science, University of Chinese Academy of Sciences, Beijing 101408, China\\
}

\date{Accepted XXX. Received YYY; in original form ZZZ}

\pubyear{\the\year{}}

\begin{document}
\label{firstpage}
\pagerange{\pageref{firstpage}--\pageref{lastpage}}
\maketitle

\begin{abstract}
This study investigates the evolution of the star-forming fraction ($F_{\mathrm{sf}}$) of Brightest Cluster Galaxies (BCGs) at $z<0.8$, using the galaxy clusters identified from the Legacy Imaging Surveys from the Dark Energy Spectroscopic Instrument (DESI). Star-forming galaxies are identified using the $g-z$ color, and $F_{\mathrm{sf}}$ is measured as a function of redshift, cluster halo mass, and galaxy stellar mass. Field galaxies are used as a comparison sample to reduce selection effects. For BCGs, $F_{\mathrm{sf}}$ increases with redshift, showing a slow rise below $z \sim 0.4 - 0.5$ and a more rapid increase above this range. In contrast, $F_{\mathrm{sf}}$ decreases with increasing cluster halo mass and BCG stellar mass. At the low stellar mass end, BCGs exhibit higher star-forming fractions than field galaxies, suggesting enhanced star formation likely fueled by cold gas accretion from the intracluster medium. Also, star-forming BCGs tend to show larger projected offsets from the optical cluster density peak than quenching BCGs, indicating ongoing assembly. The analysis of the specific star formation rate (sSFR) further indicates a transition in the dominant mechanism driving star formation in BCGs: cooling flows are likely responsible at low redshift, while gas-rich mergers play a greater role at higher redshift. The shift in dominance occurs around $z \sim 0.5$, aligning with the steep rise in $F_{\mathrm{sf}}$ of BCG.
\end{abstract}

\begin{keywords}
galaxies: clusters: general -- galaxies: evolution -- galaxies: star formation
\end{keywords}



\section{Introduction}

Brightest Cluster Galaxies (BCGs) are the most massive galaxies residing at cluster centers, characterized by their old stellar populations, high luminosity, and extended cD envelopes \citep{1987cD,2005cd, 2011BCGprofile}. Unlike typical cluster galaxies, BCGs exhibit unique properties such as Gaussian-like luminosity distributions, deviations from the Faber-Jackson relation, and distinct stellar population parameters (e.g. higher $[\alpha/\mathrm{Fe}]$ ratios), suggesting that their formation pathways differ fundamentally from other cluster galaxies \citep{1976F-Jrelation, 1995BCGlightandstdCandels, 2004BCGGaussian,  2008vel_dis, 2018alpha/Fe, 2021BCG_statistically}. The unique properties of BCGs make them valuable for a wide range of research areas, including cosmology, cluster dynamics, galaxy formation and evolution, etc. In cosmology, BCGs serve as important tracers for studying the Hubble flow, dark matter distribution, Baryon Acoustic Oscillations (BAO), and fluctuations in Cosmic Microwave Background (CMB)  \citep{1975BCG_dis, 1980dis_improve, 2009darkmatter, 2009BAO, 2009CMB, 2010hubbleflow, 2014Hubble}. Beyond their cosmological applications, understanding the formation and growth of BCGs is crucial for elucidating the physical processes driving the assembly of the most massive galaxies in the universe.

By analyzing the BCGs at $z<3$ from COSMOS2015 catalog, \cite{2019Stellar_mass_growth} identified three evolutionary phases for BCG progenitors: (1) an \textit{in situ} star formation phase characterized by gas-driven growth via internal processes at $z > 2.25$, (2) a merger and \textit{in situ} star formation phase lasting until $z \sim 1.25$, and (3) a low-redshift phase dominated by dry mergers. \cite{2015ster_mass_growth} studied BCGs in $0.2<z<1.8$ within the Spitzer Wide-Area Infrared Extragalactic (SWIRE) Survey, and demonstrated that star formation contributes negligibly to the stellar mass growth of BCGs below $z \sim 1$.

In addition to observational efforts, simulations of BCG mass growth provide critical insights. Cosmological hydrodynamical simulations by \cite{2018BCG_sim} demonstrated that stellar accretion and mergers account for more than two thirds of the total stellar mass assembled in BCGs. \cite{2018BCG_sim} further reported that the assembly redshift—the epoch at which half of the BCG's final stellar mass is assembled—decreases with increasing aperture size (the physical radius used to measure stellar mass). This suggests that BCG growth is likely an inside-out process. The findings imply that the majority of late-stage mass growth of BCGs is driven mostly by minor mergers or diffuse accretion, rather than by major mergers.  However, ongoing observational studies continue to debate the relative roles of minor and major mergers in this late-stage growth phase \citep{2015merger, 2017merger}.

At low redshifts, star formation has been shown to contribute minimally to the mass growth of BCGs, which are predominantly in quiescent and passively evolving state. Nevertheless, the hierarchical semi-analytic model proposed by \cite{2012stellar_growth_model} suggests that BCGs exhibit some star formation activity even at low redshifts ($z<0.5$). This theoretical prediction is supported by several observational studies. For example, \cite{2011BCG} analyzed 14,300 BCGs from the Sloan Digital Sky Survey (SDSS) within 0.1 < z < 0.3 and found that the fraction of blue BCGs increases from approximately 5\% in the $0.1\leq z \leq 0.2$ range to around 10\% in $0.2\leq z \leq0.3$ range, with an average of about 8\% across the entire sample. Similarly, \cite{2014BCG_sf_frac} found that 27\% of their sample of 883 BCGs at $0.09<z<0.27$ exhibit ongoing star formation, with star formation rate (SFR) up to $10 \  M_{\odot} \ \mathrm{yr}^{-1}$. Additionally, \cite{2019BCG} examined BCGs at $0.05 < z < 0.35$ and discovered that approximately 9\% of these galaxies are star-forming.

The mechanisms driving star formation in BCGs at low redshift are not yet fully understood. However, given that BCGs reside at the bottom of the cluster’s potential well, their current star formation is likely influenced by the state of the intracluster medium (ICM). 
At the cluster core, where the ICM is densest, gas from the outer regions accretes inward, losing energy through radiative cooling. This process leads to the accumulation of cooling gas in both mass and spatial extent, creating a gradual inward flow of gas toward the central galaxy. Known as the cooling flow process \citep{1994cooling}, this phenomenon suggests that BCGs should receive substantial amounts of cold gas, providing the raw material necessary to trigger star formation. \cite{2019cooling} observed the cooling flows, and \cite{2012cooling} found BCGs with star formation in cooling flow clusters typically exhibit very flat optical spectra and show the most active star formation. \cite{2016Mc} suggested that the dominant mode of fueling star formation in BCGs may have recently transitioned from galaxy–galaxy interactions in the early times to ICM cooling in the present day. However, the interplay is complicated by feedback from active galactic nuclei (AGNs), as radio-loud AGNs are overrepresented in BCGs compared to other cluster members \citep{2007AGN, 2017AGN}. AGN activity can inject energy into the ICM, counteracting cooling and thus suppressing star formation, creating a self-regulating cycle that intermittently permits or quenches star formation \citep{2024AGN_sf}. Thus, the star formation activity in low-redshift BCGs likely results from a delicate balance between cooling flows, AGN feedback, and environmental factors.

In this paper, we investigate the evolution of the star-forming fraction ($F_{\mathrm{sf}}$) in BCGs, drawn from the Data Release 9 of the Dark Energy Spectroscopic Survey
(DESI) Legacy Imaging Surveys (DESI LS-DR9). The photometric cluster catalog used in this paper, which includes over 540,000 BCGs,  has already been published in \cite{zou2019} and \cite{zou2022_dr9}. In addition to the photometric data obtained from the imaging survey, we also use the spectroscopic data from DESI Data Release 1 (DR1; \citealp{2025DESI}). Our study extends the analysis of the $F_{\mathrm{sf}}$ in BCGs to a redshift range of 0.8. This expanded sample significantly increases both the number of BCGs and the redshift range, offering a more comprehensive understanding of the BCG formation and evolution history. 

The structure of this paper is as follows. In Section \ref{sec:2}, we describe the data and sample selection. Section \ref{sec:3} focuses on the evolution of $F_{\mathrm{sf}}$ in BCGs as a function of redshift, cluster halo mass ($M_{\mathrm{200}}$), and BCG stellar mass ($M_\star$). The potential physical drivers of star formation in BCGs are discussed in Section \ref{sec:4}. A summary of the main findings and conclusions is provided in Section \ref{sec:5}. Throughout this paper, we use a flat Lambda cold dark matter ($\Lambda\mathrm{CDM}$) cosmological model, where $\Omega_{\mathrm{m}}=0.3$, $\Omega_{\Lambda}=0.7$, and $H_0=70 \mathrm{km}\ \mathrm{s}^{\mathrm{-1}} \mathrm{Mpc}^{\mathrm{-1}} $.

\section{Data and Sample Selection}\label{sec:2}
\subsection{Data Description}

\subsubsection{Photometric Data and Cluster Catalog}\label{subsec:photo data}
The DESI is a Stage IV ground-based cosmological survey aiming to explore the expansion history of the universe and the rate of structure growth with unprecedented precision, as well as to understand the nature of dark energy \citep{DESI_Collaboration_2016_a, DESI_Collaboration_2016_b, DESI_Collaboration_2024_a, 2DESI_Collaboration_2024_b}. DESI provides a vast amount of precise photometric and spectroscopic data. The photometric data are derived from the Legacy Imaging Surveys, which consist of three independent optical surveys: the Beijing-Arizona Sky Survey (BASS; \citealp{BASS;Zou.et.al.2017}), the Mayall z-band Legacy Survey (MzLS), and the DECam Legacy Survey (DECaLS), providing imaging in $g$, $r$, and $z$ band \citep{Overview_DESI_Dey.et.al.2019, zou2022_dr9}. The 5$\sigma$ depths of the optical imaging in $g$, $r$, and $z$ band are 24.7, 23.9 and 23.0 in AB mag, respectively, which are 2 magnitudes deeper than the corresponding bands in SDSS. In addition, the DESI imaging team supplements 6 years near-infrared observations from the Wide-filed Infrared Survey Explorer (WISE) in $W1$ and $W2$ bands \citep{Overview_DESI_Dey.et.al.2019, zou2022_dr9}. The 5$\sigma$ depths of the near-infrared imaging in $W1$ and $W2$ bands are 20.7 and 20.0 in AB mag, respectively. The imaging surveys provide optical and near-infrared photometric data that are used in this work. Also, these photometric data are mainly used for the target selections of the DESI spectroscopic survey. The galaxy catalog we use is sourced from the DESI LS-DR9, which was published in January 2021\footnote{\href{https://www.legacysurvey.org/dr9/}{https://www.legacysurvey.org/dr9/}}. The LS-DR9 provides  $grzW1W2$ 5-band photometry over a sky area of about 20000 deg\textsuperscript{2} in both the South and North Galactic Caps \citep{zou2022_dr9}.

\cite{zou2022_dr9} extend their previous work on estimating photo-z and detecting clusters in the DESI LS-DR9, obtaining accurate photo-z for 320 million galaxies with $r < 23$ and identifying 532,810 clusters with more than 10 members. The catalogs of the photo-z and cluster galaxies are available through the ScienceDB Web site (doi: \href{https://doi.org/10.11922/sciencedb.o00069.00003}{10.11922/sciencedb.o00069.00003}). The photo-z estimation algorithm used by \cite{zou2022_dr9} is similar to the local linear regression method in \cite{beck_2016}. This algorithm has been applied to the $ugrizW1W2$ 7-band photometric data combining SCUSS, SDSS, and WISE survey data \citep{Gao2018}, as well as to the $grzW1W2$ 5-band photometric data combining DESI and WISE survey data \citep{zou2019}. \cite{zou2022_dr9} use a new fast clustering algorithm, called Clustering by Fast Search and Find of Density Peak (CFSFDP; \citealp{Gao2020cluster}; \citealp{zou2021cluster}), to identify clusters from the selected DESI LS-DR9 galaxies. The selection criteria for the DESI LS-DR9 galaxies are:\\
    \indent1. the photo-z error is set to less than 0.1\\
    \indent2. the range of the $r$-band absolute magnitude is $-25<M_r<-16$\\
    \indent3. the stellar mass range is $6<\log M_{\star}/M_{\odot}<13$\\
    \indent4. the logarithmic mass uncertainty is less than 0.4 dex.\\
 \noindent
The CFSFDP method projects galaxies onto the 2D sky plane using photo-z, calculates local densities, and selects density peaks that are sufficiently distant from other higher-density peaks as cluster centers. Monte Carlo simulations indicate that the false-detection rate of this method is about 3.1\%. The total masses of clusters are derived using a calibrated richness–mass relation based on X-ray emission and Sunyaev–Zel’dovich effect observations. For each cluster, the BCG is identified as the $r$-band brightest galaxy within 0.5 Mpc around the density peak. Based on the detection success rate of clusters in \cite{zou2021cluster}, we retain only those clusters with $N_{\mathrm{1Mpc}} \geq 18$, where $N_{\mathrm{1Mpc}}$ is defined as the number of member galaxies within 1 Mpc around the cluster center. To create a comparison sample, we select field galaxies as those located outside the projected radius of 1 Mpc from any cluster.

\subsubsection{Spectroscopic Data}\label{subsec:spec data}
The DESI is a multi-object spectroscopic system installed at the prime focus of the 4-meter Mayall Telescope at Kitt Peak, Arizona \citep{DESI_Collaboration_2022, 2023Silber, 2024Miller, 2025DESI}. It has a 3.2° field of view, and equipped with 5,000 fiber robots that can position optical fibers on 5,000 targets and simultaneously collect their spectra. The fibers (1.5" in diameter) are connected to ten three-arm spectrographs, providing continuous wavelength coverage from 3600 to 9800 Å. The spectral resolution ranges from $R \sim 2000$ in the blue end to 5000 in the red, sufficient to resolve the [O II] doublet. The DESI spectroscopic targets include approximately 50 million extragalactic and Galactic objects, mainly from the DESI Legacy Imaging Surveys \citep{BASS;Zou.et.al.2017, Overview_DESI_Dey.et.al.2019}. During dark time, the primary observational targets include Luminous Red Galaxies (LRGs) at 0.4 < z < 1.1 \citep{Zhou2020, Zhou2023}, Emission Line Galaxies (ELGs) at 0.6 < z < 1.6 \citep{Raichoor2020, Raichoor2023AJ}, and Quasars (QSOs) at z < 3.5 \citep{Yeche2020, Chaussidon2023}.

An automated spectroscopic data pipeline has been developed to reduce raw data into wavelength- and flux-calibrated spectra, while simultaneously performing redshift calculations for all observed targets \citep{Guy2023}. The \texttt{Redrock} pipeline\footnote{\href{https://github.com/desihub/redrock}{https://github.com/desihub/redrock}} is employed for spectroscopic classification and redshift determination, which fits a suite of templates of stars, galaxies, and quasars to the DESI spectra. The spectroscopic samples we use are from the "Iron" catalog, which is released in DR1 (\citealp{2025DESI}). This catalog covers survey data collected between December 14, 2020, and June 13, 2022. The DR1 is the first major data release of the DESI collaboration, containing about ten times more spectra than the early data release in \cite{2DESI_Collaboration_2024_b}.

We use the spec-z, stellar mass and SFR data from the value-added catalog\footnote{\href{https://data.desi.lbl.gov/doc/releases/dr1/vac/stellar-mass-emline/}{https://data.desi.lbl.gov/doc/releases/dr1/vac/stellar-mass-emline/}} of stellar mass and emission line measurements (hereafter the "Iron" stellar mass catalog). This catalog is derived using the methodology described by \cite{Zou2024}, who utilized stellar population synthesis fitting based on multi-wavelength photometric data and the DESI optical spectrum. A total of 14,706,085 galaxies undergo successful stellar mass and SFR estimation. These galaxies are selected from the DR1 spec-z catalog with SPEC-TYPE="GALAXY" and ZWARN=0, indicating that they are spectroscopically confirmed galaxies with reliable redshift measurements.

\subsection{Photo-z Accuracy and Sample Redshift Selection}\label{subsec:photo-z select}
\begin{figure}
	\includegraphics[width=\columnwidth]{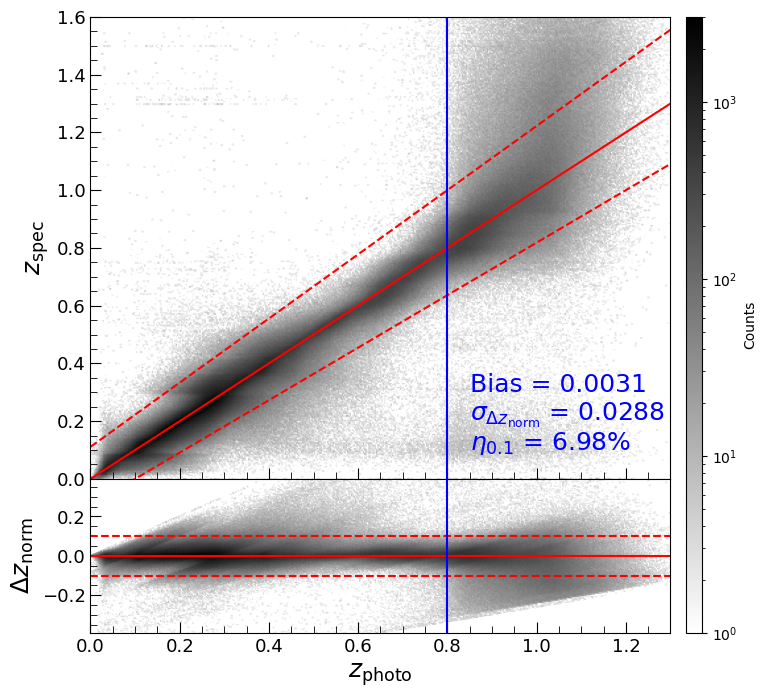}
    \caption{Top panel: the comparison between $z_{\mathrm{spec}}$ and $z_{\mathrm{photo}}$. The gray scale represents the density. The red solid line represents $z_{\mathrm{photo}} = z_{\mathrm{spec}}$. The red dashed line represents $|\Delta z_{\mathrm{norm}}| = 0.1$, and $\eta_{0.1}$ represents the fraction of galaxies outside the red dashed line. The vertical blue solid line represents the cutoff of the photo-z. The values of the three quantities (bias, dispersion and outlier rate) are marked in blue in the lower right corner.
    Bottom panel: $\Delta z_{\mathrm{norm}}$ as a function of $z_{\mathrm{photo}}$. The red solid line and dashed line represent $\Delta z_{\mathrm{norm}} = 0$ and $|\Delta z_{\mathrm{norm}}| = 0.1$, respectively.}
    \label{fig:photoz_compare}
\end{figure}

\begin{figure}
	\includegraphics[width=\columnwidth]{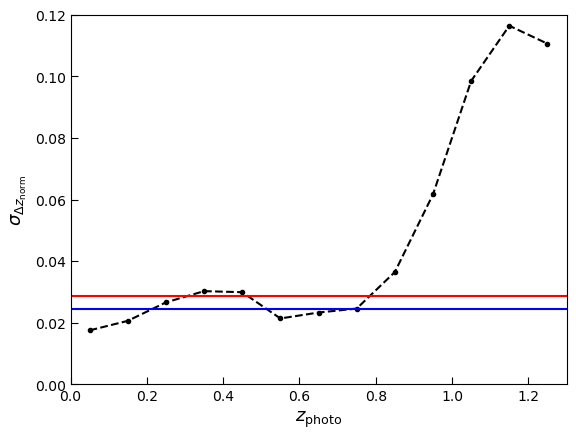}
    \caption{$\sigma_{\Delta z_{\mathrm{norm}}}$ as a function of $z_{\mathrm{photo}}$. The red solid line represents the average $\sigma_{\Delta z_{\mathrm{norm}}}$ for all galaxies that have both photo-z and spec-z, while the blue solid line represents the average $\sigma_{\Delta z_{\mathrm{norm}}}$ for the galaxies selected after applying the photo-z cut at $z_{\mathrm{photo}}<0.8$. }
    \label{fig:photoz_error}
\end{figure}

We use spec-z to assess the quality of the photo-z and to select an accurate photo-z range. The "Iron" stellar mass catalog is matched to the photo-z catalog using a matching radius of 0.5", based on the RA and Dec coordinates. Approximately 9.9 million galaxies are matched, representing about 3.1\% of the photo-z catalog. Three quantities are calculated to access the quality of photo-z estimation as follows: 
\begin{enumerate}
    \item Bias $(\overline{\Delta z_{\mathrm{norm}} })$: the median systematic deviation between the spec-z ($z_{\mathrm{spec}}$) and photo-z ($z_{\mathrm{photo}}$), after applying an iterative $3\sigma$ clipping algorithm to remove outliers. The deviation is defined as $\Delta z_{\mathrm{norm}} = (z_{\mathrm{photo}} - z_{\mathrm{spec}})/(1 + z_{\mathrm{spec}})$,  and the bias is calculated as:
    \begin{equation}
             \overline{\Delta z_{\mathrm{norm}}} = \mathrm{median} \left( \frac{z_{\mathrm{photo}} - z_{\mathrm{spec}}}{1 + z_{\mathrm{spec}}} \right).
            \label{eq:1}
    \end{equation}
    \item Dispersion ($\sigma_{\Delta z_{\mathrm{norm}}}$): 
    \begin{equation}
            \sigma_{\Delta z_{\mathrm{norm}}} = 1.48 \times \mathrm{median} \left( \frac{|\Delta z - \mathrm{median} (\Delta z)|}{1 + z_{\mathrm{spec}}} \right),
            \label{eq:2}
    \end{equation}
where $\Delta z = z_{\mathrm{photo}} - z_{\mathrm{spec}}$. With this definition, $\sigma_{\Delta z_{\mathrm{norm}}}$ is equal to the standard deviation of a
Gaussian distribution \citep{Brammer2008dispersion}. The dispersion is robust, with low sensitivity to outliers.
    \item Outlier rate ($\eta_{0.1}$): the fraction of galaxies with photo-z deviating substantially from their spec-z, and $\eta_{0.1}$ is defined as the fraction of galaxies with $|\Delta z_{\mathrm{norm}}| > 0.1$.
\end{enumerate}
The accuracy of the photo-z catalog is characterized by $\overline{\Delta z_{\mathrm{norm}}}=0.0031, \sigma_{\Delta z_{\mathrm{norm}}} = 0.0288$ and $\eta_{0.1}=6.98\%$. Figure~\ref{fig:photoz_compare} illustrates the comparison between photo-z and spec-z of the matched galaxies. A clear deviation between the two redshift estimates is observed at $z > 0.8$, where photo-z values significantly diverge from spec-z.  Figure~\ref{fig:photoz_error} shows the variation of the dispersion ($\sigma_{\Delta z_{\mathrm{norm}}}$) as a function of photo-z. Overall, $\sigma_{\Delta z_{\mathrm{norm}}}$ remains below the average value when $z_{\mathrm{photo}}<0.8$, a cutoff at $z_{\mathrm{photo}} = 0.8$ is adopted to ensure photo-z reliability. After applying this selection, the accuracy of the photo-z catalog is characterized by $\overline{\Delta z_{\mathrm{norm}}}=0.0033, \sigma_{\Delta z_{\mathrm{norm}}} = 0.0246$ and $\eta_{0.1}=3.39\%$.

\subsection{Stellar Mass} \label{subsec:stellar mass}
\begin{figure}
    \includegraphics[width=\columnwidth]{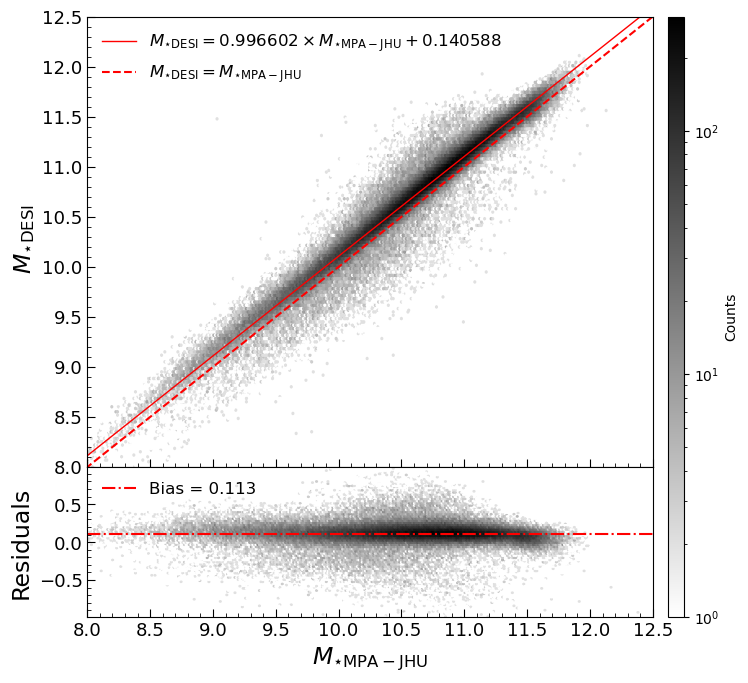}
    \caption{Top panel: the stellar mass comparison between the DESI catalog and the MPA-JHU catalog. The red dashed line represents $M_{\star\mathrm{DESI}} = M_{\star\mathrm{MPA-JHU}}$. The red solid line represents the relation of stellar mass between the two catalogs. Bottom panel: residuals as a function of $M_{\star}$. The red dashed line represents the bias of stellar mass between the two catalogs. The gray scale represents the density.}
    \label{fig:stellar_mass_compare}
\end{figure}

To obtain as many galaxies with precise stellar mass and spec-z as possible, we also incorporate stellar mass estimates in the MPA-JHU catalog\footnote{\href{https://www.sdss4.org/dr17/spectro/galaxy_mpajhu/}{https://www.sdss4.org/dr17/spectro/galaxy\_mpajhu/}} from SDSS \citep{MPA}. Notably, SDSS provides extensive observations of bright galaxies, which are not fully covered by DESI. The stellar mass in the MPA-JHU catalog is estimated using the Bayesian methodology and model grids described in \cite{2003mpa-jhu}, based on SDSS $ugriz$-band photometry with model magnitudes. The total stellar mass (lgm\_tot\footnote{\href{https://data.sdss.org/datamodel/files/SPECTRO_REDUX/galSpecExtra.html}{https://data.sdss.org/datamodel/files/SPECTRO\_REDUX/galSpecExtra.html}}) are provided. Spectroscopic data are used to correct photometric errors due to nebular emission. We match the MPA-JHU catalog with the "Iron" stellar mass catalog using a 0.5" matching radius, selecting galaxies with spec-z>0 and legal stellar mass measurements (some galaxies lack successful stellar mass measurements). A total of 319,784 galaxies are successfully matched. We compare the stellar mass estimates of these matched galaxies from both catalogs as shown in Figure~\ref{fig:stellar_mass_compare}. The relation between the stellar masses from the two catalogs is determined as: $M_{\star \mathrm{DESI}}= 0.996602 \times M_{\star \mathrm{MPA-JHU}} + 0.140588$. The stellar mass estimates from the MPA-JHU catalog show good agreement with those from the "Iron" stellar mass catalog,  with a bias of 0.113. After removing outliers, the RMS of the stellar mass differences is within 0.079 dex. The relation is used to correct the stellar masses for all galaxies in the MPA-JHU catalog, adding approximately 0.9 million galaxies and resulting in a catalog of about 15.6 million galaxies with spectroscopic redshifts.

\section{The Star-Forming Fraction of BCG}\label{sec:3}
\subsection{Identifying Star-Forming Galaxies}
We use optical color $g-z$ to distinguish whether a galaxy is star-forming or not. Optical colors can indicate the stellar populations in galaxies. In a color-color diagram, galaxies show a bimodal distribution, dividing them into a "blue cloud" of mostly star-forming systems and a red sequence of quiescent galaxies (\citealp{Strateva2001bimodal}). \cite{Williams2009bimodal} demonstrate that this bimodal distribution is seen up to $z_\mathrm{photo} \sim 2$. Therefore, we classify blue galaxies as star-forming galaxies. \cite{2011BCG} and \cite{2019BCG} use $g-r$ color to distinguish the blue BCGs, and they find the blue light is presumably due to recent star formation.

In this work, the criterion for determining whether a galaxy is blue is based on the \cite{BC03} (hereafter \hyperlink{cite.BC03}{BC03}) spectral models as discribed in \cite{2024chen}. Using the \hyperlink{cite.BC03}{BC03} templates, we establish the classification criteria of $g-z$ color as a function of redshift by tracing the galaxy template with an exponentially declining star formation history ($t=5 \mathrm{Gyr}; Z_{\odot}; \tau=2\mathrm{Gyr}$). This choice is empirically motivated: elliptical galaxies are often modeled with $\tau \sim \mathrm{1 Gyr}$ and Sb types with $\tau \sim \mathrm{5 Gyr}$, placing S0 (lenticular) systems in between, so $\tau \sim \mathrm{2 Gyr}$ is commonly adopted \citep{2000tau}; likewise, $t= \mathrm{5 Gyr}$ provides a moderate, physically plausible age for low-redshift BCG populations, yielding colours consistent with quiescent systems while retaining leverage to separate star-forming ones \citep{zou2019}. We then use this criterion to classify galaxies as red or blue. Figure~\ref{fig:g-z_photo-z_all} shows the distribution of galaxy $g-z$ color as a function of redshift for two subsamples (BCGs and field galaxies), where the solid black line represents the classification boundary. If the $g-z$ of a galaxy color is bluer (smaller) than the value of the corresponding redshift on the solid black line, we classify it as a star-forming galaxy; otherwise, it is considered as a quiescent galaxy.

\subsection{Evolution of Star-Forming Fraction of Galaxies in Different Environments} \label{subsec:Fsf_photo-z}

\begin{figure}
    \includegraphics[width=\columnwidth]{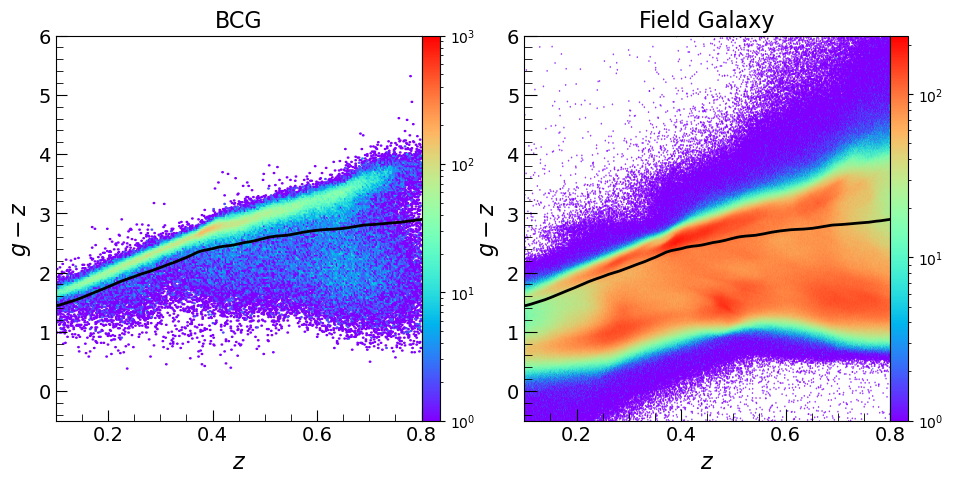}
    \caption{ $g-z$ color as a function of redshift for two subsamples. The black solid line represents the BC03 template with $\tau = 2 \mathrm{Gyr}$ \citep{BC03}. Galaxies below the solid black line are classfied as star-forming galaxies. The color scale represents the density. }
    \label{fig:g-z_photo-z_all}
\end{figure}

We impose a cutoff of $z_{\mathrm{photo}}>0.1$ due to the small sample size of BCGs in the limited comoving volume at $z_{\mathrm{photo}}<0.1$. The same redshift range is applied to field galaxies to maintain consistency in the analysis. The two subsamples include 221,256 BCGs and 188,763,669 field galaxies in the photo-z range of 0.1-0.8. The distribution of galaxy colors in this redshift range is shown in Figure~\ref{fig:g-z_photo-z_all}. As seen in Figure~\ref{fig:g-z_photo-z_all}, a significant portion of field galaxies are star-forming, while BCGs are predominantly composed of quiescent galaxies.

\begin{figure}
    \centering
    \includegraphics[width=\columnwidth]{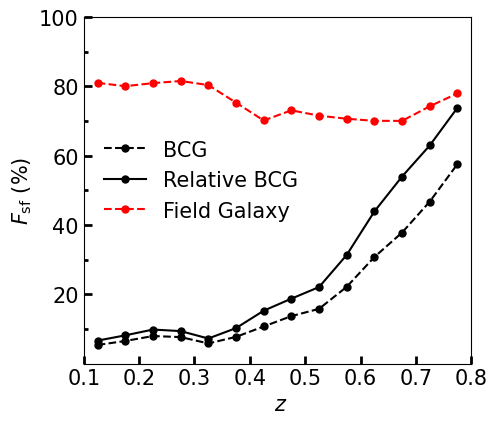}
    \caption{The evolution of the $F_{\mathrm{sf}}$ as a function of redshift for BCGs and field galaxies. The solid line represents the relative $F_{\mathrm{sf}}$ for BCGs. The Poisson errors are too small to be ignored.}
    \label{fig:sfgf-all}
\end{figure}

We calculate the overall star-forming fraction ($F_{\mathrm{sf}}$) for the two subsamples, with overall values of 16.97\% for BCGs and 74.51\% for field galaxies. \cite{2011BCG} and \cite{2019BCG} focused on BCG samples with photo-z less than 0.35, resulting in a lower overall fraction of star-forming BCGs compared to our study. To enable a direct comparison, we restrict BCGs to the same redshift range ($z_{\mathrm{photo}}<0.35$), and find that the fraction of star-forming BCGs is approximately 7\%, which is consistent with the $F_{\mathrm{sf}}$ of BCG in \cite{2011BCG} and \cite{2019BCG}. Furthermore, we analyze the evolution of the $F_{\mathrm{sf}}$ with redshift, as shown in the 
Figure~\ref{fig:sfgf-all} (see the dashed lines). Uncertainties in $F_{\mathrm{sf}}$ are estimated assuming Poisson counting statistics for the numbers of BCGs and star-forming BCGs; given our large sample sizes, the resulting errors are negligible and are not shown. For BCGs, $F_{\mathrm{sf}}$ increases with redshift, showing a slow rise below $z \sim 0.4-0.5$ and more rapid increase above $z \sim 0.4-0.5$. The observed increase in $F_{\mathrm{sf}}$ in BCGs with redshift is consistent with \cite{2011BCG} and \cite{2019BCG}. Additionally, \cite{2015ster_mass_growth} also reported an increase in star formation activity with redshift in BCGs. In contrast, for field galaxies, $F_{\mathrm{sf}}$ shows a slight decline around $z \sim 0.5-0.6$, followed by a rise at higher redshifts. However, these trends for field galaxies may be strongly influenced by sample selection effects. At low redshifts, the observed population includes numerous low-mass galaxies in early evolutionary stages, while high-redshift samples are dominated by relatively massive galaxies due to Malmquist bias \citep{1925select——effect, 2015GalaxyFormation}. The latter are more likely to be quenched systems, potentially biasing the observed $F_{\mathrm{sf}}$ evolution. We also observe an small dip in $F_{\mathrm{sf}}$ for field galaxies around $z \sim 0.4$. This is attributed to the inaccuracies in photo-z, as discussed in Section \ref{subsec:photo-z select}. Overall, in comparison to field galaxies, BCGs have different evolutionary processes, and therefore, their $F_{\mathrm{sf}}$ is generally lower and the evolution of $F_{\mathrm{sf}}$ with redshift shows distinct characteristics. 

To reduce the impact of selection effects and better characterize the intrinsic $F_{\mathrm{sf}}$ of BCGs, we calculate the relative $F_{\mathrm{sf}}$. The relative $F_{\mathrm{sf}}$ is defined as the $F_{\mathrm{sf}}$ normalized by the $F_{\mathrm{sf}}$ of field galaxies, i.e., the $F_{\mathrm{sf}}$ of BCGs divided by that of field galaxies. The solid line of Figure~\ref{fig:sfgf-all} shows the evolution of the relative $F_{\mathrm{sf}}$ with redshift for BCGs. The relative $F_{\mathrm{sf}}$ for both BCGs shows a notably stronger increasing trend. This indicates that, compared to field galaxies, 
the star-forming activity of BCGs becomes stronger with increasing redshift. Notably, the relative $F_{\mathrm{sf}}$ of BCGs also displays different growth rates before and after redshift 0.4, suggesting distinct driving mechanisms for star formation activity.

\subsection{Evolution of Star-Forming Fraction in Different Halo Mass}\label{subsec: sfgf-different-halo}

\begin{figure}
    \includegraphics[width=\columnwidth]{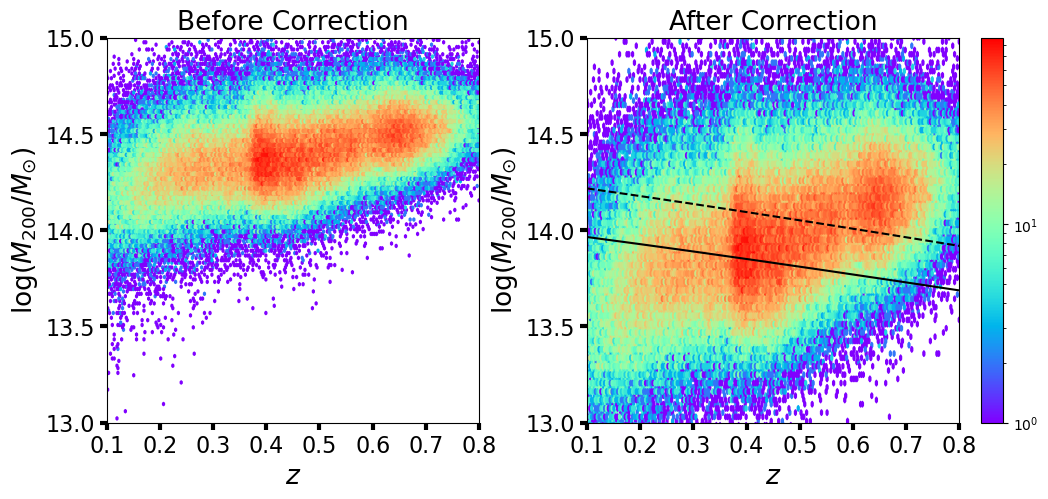}
    \caption{Left panel: Distribution of uncorrected dark matter halo mass $M_\mathrm{200}$ with BCG photo-z. 
    Right panel: Distribution of corrected dark matter halo mass $M_\mathrm{200}$ with BCG photo-z. The black solid line represents the accretion history for a halo with $M_{\star} = 1\times10^{\mathrm{14}}M_\odot$ in $z=0$, the black dashed line represents the accretion history for a halo with mass $M_{\star} = 2\times10^{\mathrm{14}}M_\odot$ in $z=0$. The mass accretion history of halos is based on the extended Press–Schechter formalism in \protect \cite{commah}. The lines correspond to the boundaries of the three cluster subsets. The color scale represents the density. Briefly, around $z\sim0.4$ photo-z uncertainties are larger than in adjacent bins, which can modestly accentuate the apparent density rise just below this redshift.}
    \label{fig:M200}
\end{figure}

In order to investigate the effect of cluster halo mass on the $F_{\mathrm{sf}}$ of BCGs, we divide the cluster samples into three subsets based on the mass of their host halos. \cite{zou2021cluster} 
provided the halo masses of $M_{500}$ for their clusters, based on the calibration related to the total member luminosity ($L_{\mathrm{1 \ Mpc}}$). We use the \texttt{colossus}\footnote{\href{https://bdiemer.bitbucket.io/colossus/halo\_mass\_defs.html}{https://bdiemer.bitbucket.io/colossus/halo\_mass\_defs.html}} package to convert $M_{500}$ to $M_{200}$. When obtaining the empirical calibration relation for $M_{500}$, \cite{zou2021cluster} used the data associated with high halo masses (as shown in Figure 3 in \citealp{zou2021cluster}). As a result, the $M_{500}$ estimates, as well as the converted $M_{200}$, tend to show greater inaccuracies in the lower halo mass range. \cite{chen2025} measures halo masses using the weak lensing method and provides a correction between \( M_{200} \) measured by lensing and the \( M_{200} \) we used:
\begin{equation}
        \log_{10}M_{\mathrm{200,lens}} = 1.71 \times \log_{10}M_{\mathrm{200,our}} - 10.29.
\end{equation}
Figure~\ref{fig:M200} shows $M_{200}$ as a function of redshift before and after the correction. In this figure, we can see that the lensing method primarily corrects the low-mass halos. After correction, the halo masses of the cluster samples are predominantly above $10^{13} M_\odot$.

To define three subsets with different halo masses, we divide the samples based on dark matter halos with $M_{\mathrm{200}}=1 \times 10^{\mathrm{14}}M_{\odot}$ and $M_{\mathrm{200}}=2 \times 10^{\mathrm{14}}M_{\odot}$ at $z=0$. We trace their evolution back to $z_{\mathrm{photo}}=0.8$ to reconstruct the accretion histories, using the \texttt{commah}\footnote{\href{https://github.com/astroduff/commah}{https://github.com/astroduff/commah}} package, which calculates the mass accretion history of halos based on the extended Press–Schechter formalism \citep{commah, 2015commah, 2015commah1}. The dashed and solid lines in the right panel of Figure~\ref{fig:M200} illustrate the evolution of these two reference halo masses. Accordingly, we divide the cluster samples into low-mass halo clusters ($1 \times 10^{\mathrm{13}}M_{\odot} < M_{\mathrm{200}} < 1 \times 10^{\mathrm{14}}M_{\odot}$ at $z=0$), intermediate-mass halo clusters ($1 \times 10^{\mathrm{14}}M_{\odot} \leq M_{\mathrm{200}} < 2 \times 10^{\mathrm{14}}M_{\odot}$ at $z=0$) and high-mass halo clusters ($M_{\mathrm{200}} \geq 2 \times 10^{\mathrm{14}}M_{\odot}$ at $z=0$).

\begin{figure}
	\includegraphics[width=\columnwidth]{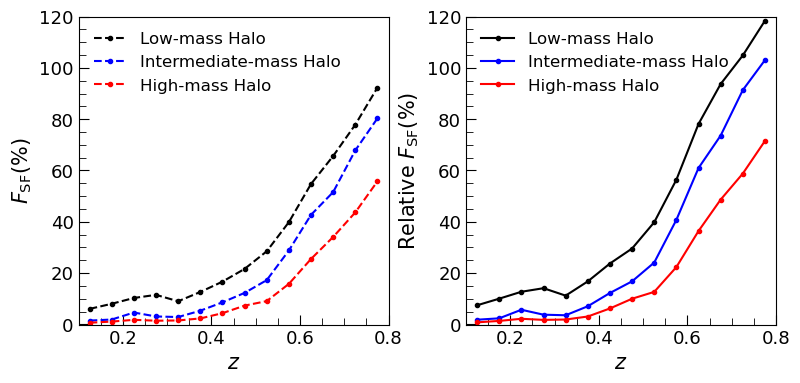}
    \caption{The $F_{\mathrm{sf}}$ and the relative $F_{\mathrm{sf}}$ of BCG as a function of redshift in different halo masses. The Poisson errors are too small to be ignored.}
    \label{fig:SFGF-halomass}
\end{figure}

We present the $F_{\mathrm{sf}}$ for BCGs as a function of redshift in different halo mass ranges in the left panel of Figure~\ref{fig:SFGF-halomass}. For BCGs, the $F_{\mathrm{sf}}$ at a specific redshift decreases with the increasing halo mass. The negative correlation between the $F_{\mathrm{sf}}$ of BCGs and halo mass is consistent with the conclusion of \cite{2019BCG}. \cite{2019BCG} focused on samples within the photo-z range of $0.05 \leq z < 
0.35$, with halo masses concentrated between $6 \times 10^{13} M_\odot$ and $10^{14.3} M_\odot$. In comparison, our analysis extends to a broader range of both redshift and halo mass, covering $0.1<z<0.8$ and $1 \times 10^{\mathrm{13}}M_{\odot} < M_{\mathrm{200}} < 10^{\mathrm{14.7}}M_{\odot}$. The evolution of the $F_{\mathrm{sf}}$ with redshift in BCGs across different halo mass ranges is consistent with the findings in Section \ref{subsec:Fsf_photo-z}. When the redshift is less than 0.4-0.5, the $F_{\mathrm{sf}}$ of BCGs in three halo mass ranges remains low. In high-mass halos, the $F_{\mathrm{sf}}$ is close to zero. In intermediate-mass halos, the $F_{\mathrm{sf}}$ of BCGs is around 5\% and remains nearly constant with increasing redshift. In low-mass halos, the average $F_{\mathrm{sf}}$ of BCGs is around 10\% and shows an slight increasing trend with redshift. As the redshift exceeds 0.4-0.5, the $F_{\mathrm{sf}}$ of BCGs in all three halo mass ranges increases rapidly, with the faster growth occurring in lower-mass halos. The average $F_{\mathrm{sf}}$ of BCGs at $z>0.4$ in low-mass halos is 1.3 times higher than that in intermediate-mass halos and 2.2 times higher than that in high-mass halos, while the $F_{\mathrm{sf}}$ in intermediate-mass halos is 1.7 times higher than that in high-mass halos. These comparisons suggest that BCGs in low-mass halos exhibit more active star formation.

The right panel of Figure~\ref{fig:SFGF-halomass} shows the evolution of the relative $F_{\mathrm{sf}}$ of BCGs, using the field galaxies in Figure~\ref{fig:sfgf-all} as a reference, as a function of redshift in different halo mass ranges. The relative $F_{\mathrm{sf}}$ of BCGs increases with redshift across all halo mass ranges. In low-mass halos at $z>0.7$, the relative $F_{\mathrm{sf}}$ of BCGs exceeds 100\%, indicating higher $F_{\mathrm{sf}}$ compared to field galaxies. This may be because, at high redshifts, BCGs in such halos are still in early evolutionary stages, where AGN activity is not yet dominant, allowing efficient accretion of cold gas \citep{2024AGN_sf}. BCGs in intermediate- and high-mass halos exhibit relative $F_{\mathrm{sf}}$ below 100\% over the redshift range $0.1 < z < 0.8$, indicating lower $F_{\mathrm{sf}}$ compared to field galaxies. This suppression is likely due to the dense environments of these halos, where gravitational interactions along with AGN activity, significantly suppress their star formation \citep{1996GalaxyHarassment, 2006dense_envir, 2024AGN_sf}.

\subsection{Evolution of Star-Forming Fraction with Stellar Mass}\label{subsection:SFGF-stellarmass}

To ensure accurate stellar mass measurements, we use BCGs with available spec-z and stellar masses derived using spectroscopic information in the "Iron" stellar mass catalog. These measurements are consistent with those from the MPA-JHU catalog, ensuring the reliability of our analysis. In total, 55,765 BCGs with precise spec-z and stellar mass are included. Based on these BCGs, we analyze the evolution of $F_{\mathrm{sf}}$ as a function of stellar mass within the redshift range $0.1<z<0.8$ and the stellar mass range $10^{10.7}M_{\odot}<M_{\star}<10^{12.1} M_{\odot}$. Previous studies \citep{2015halo,2016halo, 2019halo} have found a positive correlation between BCG stellar mass and cluster halo mass (see this correlation for our sample in APPENDIX \ref{appendix A}). Given this correlation, we expect the $F_{\mathrm{sf}}$ for BCGs to decrease with increasing stellar mass.

\begin{figure}
    \includegraphics[width=\columnwidth]{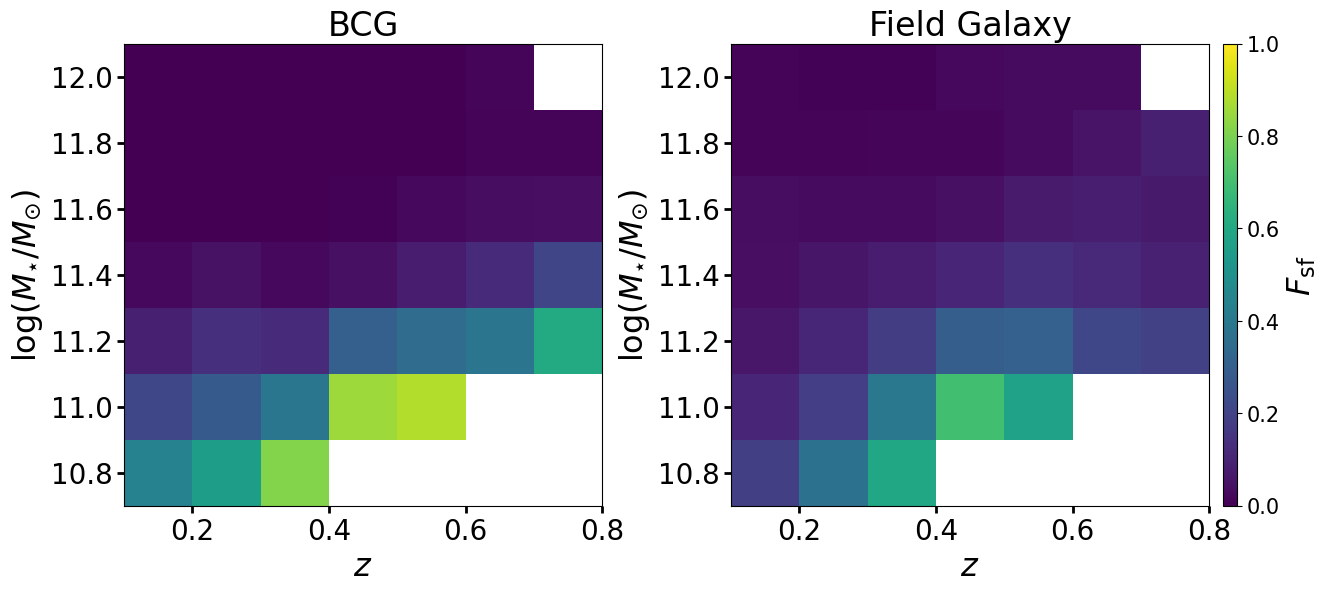}
    \caption{The evolution of $F_{\mathrm{sf}}$ as a function of redshift and stellar mass for BCGs and field galaxies. Galaxies are grouped into redshift bins with an interval of 0.1, and further subdivided into stellar mass intervals of 0.2 dex. Each resulting grid cell, defined by a specific redshift and stellar mass range, is used to calculate $F_{\mathrm{sf}}$. Cells with fewer than 100 galaxies are excluded from the analysis. The color scale represents $F_{\mathrm{sf}}$.}
    \label{fig:ABS_Frac_stellar}
\end{figure}

To compare BCGs with field galaxies, we construct comparison samples by selecting galaxies that match the redshift and stellar mass distribution of BCGs. Specifically, for each BCG, we select one field galaxy that are best matched in redshift (within 0.02) and stellar mass (within 0.05 dex). This one-to-one matching ensures that the two samples have comparable distributions in redshift and stellar mass. We calculate $F_{\mathrm{sf}}$ in each redshift-$M_\star$ grid cell for the two samples. The evolution of $F_{\mathrm{sf}}$ as a function of stellar mass and redshift is shown in Figure~\ref{fig:ABS_Frac_stellar}. To ensure statistical reliability, cells containing fewer than 100 galaxies are excluded from the calculation. Our results show that $F_{\mathrm{sf}}$ decreases with increasing stellar mass for both samples, with the evolution of BCG $F_{\mathrm{sf}}$ consistent with the findings of \cite{2022BCG_important}. Additionally, the $F_{\mathrm{sf}}$ for BCGs increases with redshift, in agreement with previous results. However, there are stronger selection effects in spectroscopic measurements, along with the limited number of high-redshift spectroscopic samples for field galaxies. This is particularly evident at the low-mass end, where the $F_{\mathrm{sf}}$ for field galaxies in spec-z first increases and then decreases.

\begin{figure}
    \includegraphics[width=\columnwidth]{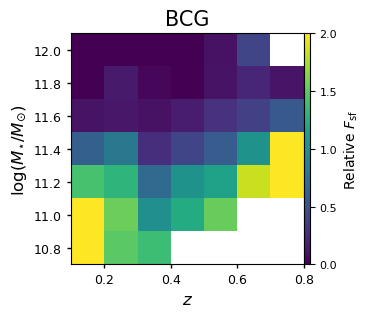}
    \centering
    \caption{The evolution of relative $F_{\mathrm{sf}}$ as a function of redshift and stellar mass for BCGs. Cells with fewer than 100 galaxies are excluded from the analysis. The color gradient represents
    the relative $F_{\mathrm{sf}}$.}
    \label{fig:Rela_Frac_stellar}
\end{figure}

To mitigate the selection effects, field galaxies are used as a control sample to calculate the relative $F_{\mathrm{sf}}$ for BCGs. Figure~\ref{fig:Rela_Frac_stellar} illustrates the evolution of the relative $F_{\mathrm{sf}}$ as a function of redshift and stellar mass. In the left panel of Figure~\ref{fig:Rela_Frac_stellar}, the yellow-bright cells, representing higher relative $F_{\mathrm{sf}}$ values (around 2), are concentrated in the low-mass end of BCGs. This indicates that low-mass BCGs exhibit much higher $F_{\mathrm{sf}}$ compared to field galaxies. Specifically, for $z<0.3$, BCGs with stellar mass below $10^{11.2} M_{\odot}$ exhibit a relative $F_{\mathrm{sf}}$ between 1.7 and 2.5. At higher redshift of $0.3<z<0.6$, BCGs with below $10^{11.2} M_{\odot}$ have a relative $F_{\mathrm{sf}}$ of approximately $1.3 \sim 1.5$ and for $z>0.6$, BCGs with stellar mass below $10^{11.4} M_{\odot}$ show a relative $F_{\mathrm{sf}}$ between 1.7 and 2.5. At the high-mass end, the $F_{\mathrm{sf}}$ of BCGs is nearly zero, while that of field galaxies is also low, typically ranging from 1\% to 10\%. In cases where BCGs have non-zero $F_{\mathrm{sf}}$, the $F_{\mathrm{sf}}$ of field galaxies is approximately ten times higher than that of BCGs.

\subsection{$F_{\mathrm{sf}}$ of BCGs versus Projected Offset}\label{subsection:SFGF-offset}

\begin{figure}
    \includegraphics[width=\columnwidth]{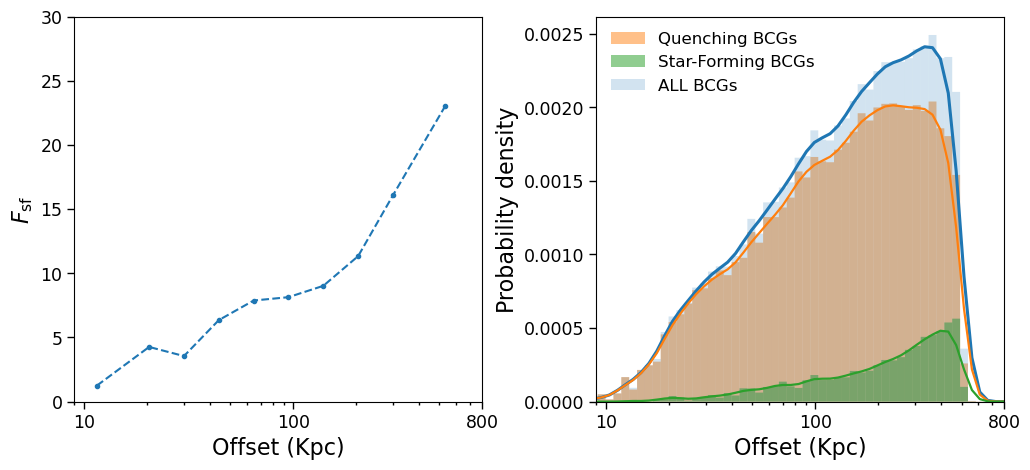}
    \centering
    \caption{Left panel: The evolution of $F_{\rm sf}$ as a function of projected offset. Right panel: Kernel-smoothed probability density estimates for the distribution of projected offsets.}
    \label{fig:offset}
\end{figure}

Beyond the dependencies on redshift, cluster halo mass, and galaxy stellar mass, BCG star-forming activity is also tightly linked to the projected offset between the BCG and the cluster centre. Here the cluster center is taken to be the density peak in the \cite{zou2022_dr9} catalog. Following \citet{2009offset}, we measure the projected BCG–centre offset for our optical sample. To conservatively ensure that the BCG and the density peak trace the same cluster, we impose a photo-z–based consistency cut: the velocity difference, defined as $v=c\,[z_{\rm phot,peak}-z_{\rm phot,BCG}]/(1+z_{\rm phot,BCG})$, should satisfy $|v|\le 4000\,\mathrm{km\,s^{-1}}$\citep{Wojtak2011Nature}. We compute the projected offset as the separation between the BCG and the optical cluster density peak. The left panel of Figure~\ref{fig:offset} shows the evolution of $F_{\mathrm{sf}}$ as a function of offset, $F_{\mathrm{sf}}$ increases with increasing offset and reaches a maximum at around 400 kpc. The right panel of Figure~\ref{fig:offset} presents kernel-smoothed probability density estimates of the projected BCG–centre offsets, comparing star-forming and quenching BCGs. We find that star-forming BCGs are more displaced from the centre, whereas quenching BCGs lie closer to the centre. This implies that star-forming BCGs are subject to stronger interactions with other galaxies (e.g., major mergers), in agreement with \cite{2021BCG_statistically}.

\cite{2021BCG_statistically} found that when the BCG has a large offset ($>\mathrm{0.2 Mpc/h}$), its stellar mass is consistent with the upper limit expected from the member galaxy mass distribution—meaning it can be explained by normal statistics. However, for small-offset BCGs ($\leq \mathrm{0.2 Mpc/h}$), the stellar mass exceeds this expected maximum, suggesting additional central assembly or a particularly enhanced growth process. They also found that over half of clusters with BCG offsets larger than $\mathrm{0.2 Mpc/h}$ host a second-ranked galaxy that is likewise offset by $>\mathrm{0.2 Mpc/h}$. This finding suggests that many of these systems are in the process of merging, with the original BCGs from the progenitor clusters still displaced from the centre of the newly forming potential well. These trends are consistent with our finding that star-forming BCGs preferentially inhabit the large-offset regime, with our star-forming BCG offsets concentrated around $300\text{--}400 \mathrm{kpc}$.

Taken together, large BCG–centre offsets signal recent halo mergers or ongoing assembly, while small offsets indicate earlier assembly and relaxed systems. This picture is also consistent with \cite{2014offset}, whose N-body simulations with semi-analytic treatments of galaxy formation, merging, and tidal disruption show that major and semi-major cluster mergers can displace BCGs from the potential minimum.

\section{Discussion}\label{sec:4}

\subsection{Reliability of Color-Based Star-Forming Galaxy Identification}

\cite{2019BCG} found that while quiescent BCGs predominantly exhibit red optical colors, star-forming BCGs can display both blue and red colors, suggesting that star formation may be obscured by dust in some cases. \cite{2019BCG} used $g-r$ optical colors and WISE infrared data to distinguish distinguish between star forming and non-star forming. In contrast, we use $g-z$ colors, as the $ z $ band falls in a longer wavelength range, making it less susceptible to dust extinction than the $ r $ band. We further assess the reliability of using $ g - z $ colors as a criterion for identifying star-forming galaxies.

\cite{2012bimodality} found that all galaxies, regardless of whether they are classified as central or satellite, exhibit a similar bimodal distribution of specific star formation rate (sSFR), with a strong break at $\mathrm{sSFR} \sim 10^{-11} \mathrm{yr}^{-1}$. To assess the reliability of color-based star-forming galaxy identification, we utilize sSFR measurements derived from SED fitting performed with the CIGALE code, as presented in \cite{Zou2024}. There are 7.3 million galaxies in total have reliable sSFR data in the redshift range of 0.1 to 0.8. The reason we do not initially use the sSFR data to identify star-forming galaxies is that the number of galaxies with reliable sSFR measurements is significantly lower than that of photometric galaxies. For example, in the case of BCGs, only 18\% of those with photometric data have reliable sSFR measurements.   

If a galaxy’s sSFR is larger than $10^{-11} \mathrm{yr}^{-1}$, we consider it to be a true star-forming galaxy. We verify the reliability of color-based star-forming galaxy identification from two perspectives: first, by checking whether galaxies with sSFR greater than $10^{-11} \mathrm{yr}^{-1}$ are classified as blue galaxies (star-forming galaxies) based on their $g-z$ color; second, by checking whether blue galaxies have an sSFR greater than $10^{-11} \mathrm{yr}^{-1}$. These two perspectives allow us to assess the completeness and purity of the color-based star-forming galaxy identification.

Among galaxies with sSFR greater than  $10^{-11} \mathrm{yr}^{-1}$, 88.66\% are classified as blue galaxies based on their $g-z$ color. Meanwhile, among galaxies classified as blue galaxies, 92.18\% are identified as star-forming galaxies based on their sSFR. The high completeness suggests that the majority of actual star-forming galaxies are successfully included in the $g-z$ color selection, while the high purity confirms that most selected blue galaxies are truly star-forming. Furthermore, across all galaxies, the $F_{\mathrm{sf}}$ derived from the sSFR criterion is 54.54\%, while the $F_{\mathrm{sf}}$ derived from the  $g - z$  color criterion is 52.46\%. The close agreement between the star-forming BCG fraction derived from the $g-z$ color selection and that from the sSFR classification both supports the reliability of our BC03 parameter choice and confirms that g-z colour is an effective selector of star-forming galaxies in our samples.

\subsection{Enhanced \texorpdfstring{$F_{\mathrm{sf}}$}{Fsf} for Low-mass BCGs Compared with Field Galaxies} \label{subsec: ssfr}

\begin{figure}
	\includegraphics[width=\columnwidth]{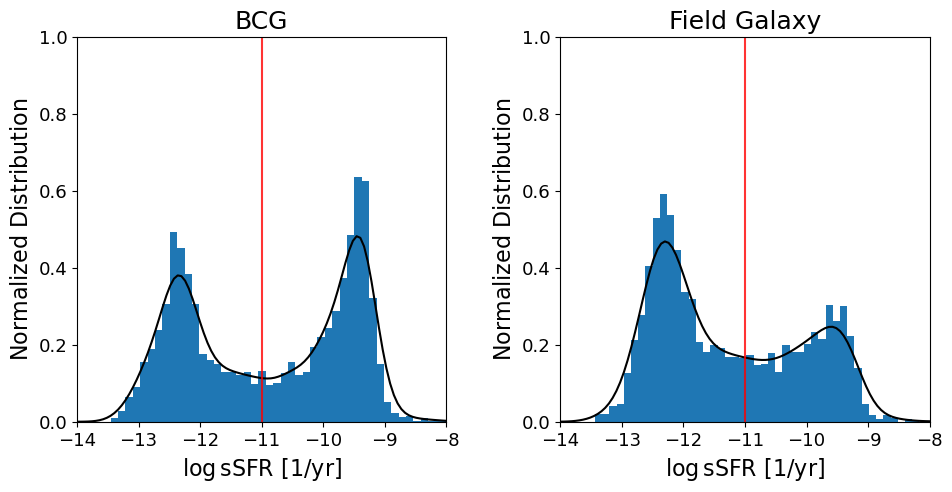}
    \caption{Normalized probability density distribution of sSFR for low-mass BCGs and field galaxies with $10^{10.7} M_{\odot} < M_{\star} < 10^{11.3} M_{\odot}$. The red solid line represents the sSFR threshold of $10^{-11} [1/\mathrm{yr}]$ to distinguish star-forming galaxies \citep{2012bimodality}.}
    \label{fig:sSFR_dis_lowmass}
\end{figure}

In Section \ref{subsection:SFGF-stellarmass}, we find that BCGs exhibit a higher star-forming
galaxy fraction at the low mass end compared to field galaxies. We use sSFR to verify this conlusion and present the evolution of the sSFR of low-mass BCGs and field galaxies as a function of redshift. Low-mass BCGs and field galaxies are selected with $M_{\star} < 10^{11.3} M_{\odot}$, where the relative $F_{\mathrm{sf}}$ is significantly high (approximately above 1.5). The sSFR distribution for the low-mass BCGs and field galaxies are shown in Figure~\ref{fig:sSFR_dis_lowmass}. The sSFR distributions for both populations exhibits bimodality. For BCGs, the region with $\log \mathrm{sSFR}>-11$ slightly dominates, with approximately 55\% classified as star-forming.  In contrast, for field galaxies, the region with $\log \mathrm{sSFR}<-11$ is more prominent, indicating a dominant quiescent population, with only about 39\% of galaxies being star-forming.

\begin{figure}
	\includegraphics[width=\columnwidth]{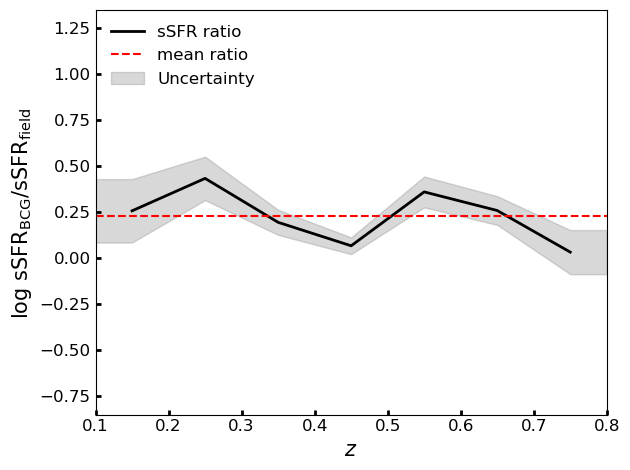}
    \caption{The evolution of sSFR ratio with redshift for low-mass BCGs and field galaxies with $10^{10.7} M_{\odot} < M_{\star} < 10^{11.3} M_{\odot}$. The black solid line shows the evolution of the sSFR ratio, with the shaded region indicating the uncertainty range. The red dashed line represents the average sSFR ratio, which is 0.23 dex.}
    \label{fig:lowmassBCG_sSFR}
\end{figure}

Furthermore, we calculate the redshift evolution of the sSFR ratio ($\log\mathrm{sSFR}_{\mathrm{BCG}} / \mathrm{sSFR}_{\mathrm{field}}$) between low-mass BCGs and field galaxies. The redshift range is divided into bins with intervals of 0.1. In each redshift bin, the combined sSFR is calculated as the sum of individual SFRs divided by the sum of stellar masses, following the approach of \cite{2016Mc}. Figure~\ref{fig:lowmassBCG_sSFR} shows the evolution of sSFR ratio with redshift for low-mass BCGs and field galaxies in the same mass range. Notably, the sSFR of low-mass BCGs is on average 0.23 dex higher than that of field galaxies across all redshift ranges. This analysis indicates that low-mass BCGs exhibit more active star formation than field galaxies with similar stellar mass.

\subsection{Possible Star Formation Mechanisms of BCGs at \texorpdfstring{$z<0.8$}{z<0.8}}\label{subsection:sf_mechanisms}

In Section \ref{sec:3}, we find that the growth rate of $F_{\mathrm{sf}}$ in BCGs differs between high and low redshifts. This difference might be related to the mechanisms driving the star formation of BCGs. According to \cite{2016Mc}, BCGs undergo two distinct phases of star formation: for $z<0.6$, star formation is predominantly triggered by cooling flows, whereas for $z>0.6$, it is predominantly primarily driven by gas-rich mergers. Based on a sample of 90 BCGs from the South Pole Telescope, they calculated the sSFR and studied its evolution within the redshift range of 0.25–1.25. This evolution can be described as the sum of two exponential functions in time ($\langle\mathrm{sSFR}\rangle \propto e^{t/\tau}$), with the best-fitting models having decay times of 4 Gyr (low-z) and 0.7 Gyr (high-z). We also check the sSFR evolution with our spectroscopic BCG samples at $z < 0.8$. 

\begin{figure}
	\includegraphics[width=\columnwidth]{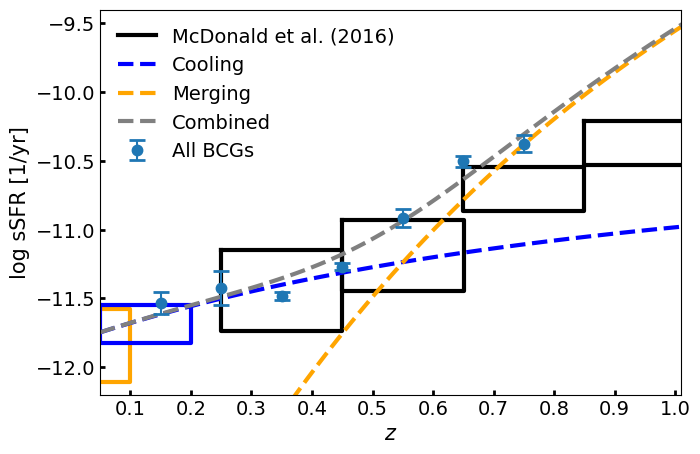}
    \caption{The evolution of sSFR with redshift for all BCGs.  The blue points represent the combined sSFR results for our entire BCG sample. The blue dashed line represents the cooling flow component (decay time of 4 Gyr), the yellow dashed line represents the merger component (decay time of 0.6 Gyr), and the gray dashed line illustrates the combined model including both components. The boxes represent the combined sSFR across different redshift bins, with the vertical size of the boxes indicating the associated uncertainties \protect \citep{2016Mc}. The yellow and blue boxes represent nearby galaxy clusters, with data sourced from \protect \cite{2010sSFRfig1} and \protect \cite{2014sSFRfig2}.}
    \label{fig:sSFR_evolution}
\end{figure}

Figure~\ref{fig:sSFR_evolution} shows the evolution of the sSFR with redshift for our BCGs. The blue points indicate the combined sSFR of our BCG sample in different redshift bins. At redshift below 0.5, the sSFR shows a relatively gradual increase with redshift, while at redshift above 0.5, the sSFR exhibits a significant upward trend. This apparent change in trend is captured by a two-component exponential fit: the blue dashed line corresponds to a slowly decaying component with a timescale of 4 Gyr, and the yellow dashed line represents a more rapidly decaying component with a timescale of 0.6 Gyr. The gray dashed line illustrates the sum of these two components, representing a two-phase evolutionary model that incorporates both cooling flow and merger-driven star formation. This two-phase model suggests that star formation in BCGs is predominantly fueled by cooling flows at low redshifts and by gas-rich mergers at higher redshifts. The transition between these two dominant mechanisms occurs around $z \sim 0.5$, whereas in \cite{2016Mc} the transition is found at a slightly higher redshift of $z \sim 0.6$. This transition redshift also coincides with the behavior of the $F_{\mathrm{sf}}$, which increases gradually below $z \sim 0.4{-}0.5$, followed by a more rapid rise above this redshift. This consistency between sSFR evolution and $F_{\mathrm{sf}}$ supports the interpretation of a change in the dominant star formation mechanism. Compared to the results of \cite{2016Mc}, our fitted merger component shows a slightly steeper decline, with a decay time of 0.6 Gyr, versus 0.7 Gyr in their study. At redshifts z < 0.5, our sSFR measurements are in good agreement with their results. However, at z > 0.5, our BCGs show systematically higher sSFR values. This discrepancy may partly arise from the limited number of high-redshift BCGs in \cite{2016Mc} sample, which included only 22 BCGs in the range 0.6 < z < 0.8, and just 14 BCGs at z > 0.8, leading to larger statistical uncertainties. Nonetheless, the deviation remains within $3\sigma$.

\subsection{Cooling Flows as Triggers for Star Formation at Low Redshift }

In \cite{2016Mc}, low-redshift star formation activity is attributed to the replenishing fuel from the cooling ICM in relaxed, cool core clusters.  Studies have already observed the presence of cooling flows and confirmed their role in promoting star formation activity \citep{2017cool, 2018cooling, 2019cooling}. Hot gas can release energy by emitting X-ray photons, cooling down and forming cooling flows. These cooling flows move toward the center of the cluster, providing a continuous supply of cold gas to the BCG. The deposition of this cold gas can stimulate star formation activity in the central region of the galaxy.

We suggest that the stronger star-forming activity in low-mass ($10^{10.7} M_{\odot} \sim 10^{11.1} M_{\odot}$) and low-redshift BCGs relative to field galaxies could be due to the supply of cold gas from cooling flows. This interpretation is consistent with the low redshift regime (z < 0.2), where \textit{in situ} star formation—fed by cooling flows—has been shown to dominate BCG mass assembly in cosmological simulations \citep{2018BCG_sim}. At low redshifts, low-mass BCGs are often located in dynamically young clusters that are still undergoing assembly. In such environments, the thermal energy might be insufficient to completely offset radiative cooling, particularly in the dense central regions of the cluster core. As a result, cooling flows may develop, channelling cold gas inward along the gravitational potential well of the forming cluster. The BCG, located at the center of this well, could potentially accrete some fraction of this inflowing cold gas, which may then contribute to enhanced in-situ star formation \citep{2011BCG, 2024AGN_sf}. Take Abell 1835 for example, this cluster has been observed in the DESI spectroscopic survey and also features multi-wavelength data, including X-ray and millimeter-wave observations. \cite{2019cooling} presented Atacama Large Millimeter/submillimeter Array (ALMA) observations of Abell 1835, detecting molecular gas in the cool core of the cluster, as shown in Fig.3 of their paper. The BCG presents stellar mass of $1.6 \times 10^{10} M_{\odot}$, and an sSFR of $10^{-8} \mathrm{yr}^{-1}$, indicating high star-forming activity. The detected cold gas provides the necessary fuel for this ongoing star formation. It has been suggested that cold gas in cluster centers primarily originates from localized radiative cooling of the intracluster medium (ICM), particularly under the combined influence of AGN-driven bubble uplift, thermal instabilities, and turbulence \citep{2019cooling}. These processes are thought to jointly promote the formation of multiphase, filamentary structures that condense out of the hot ICM and subsequently precipitate toward the cluster core. Such cold filaments can then serve as a fuel reservoir for star formation in BCG.

Low-redshift BCGs are more readily accessible for multi-wavelength observations, which allow for more precise constraints on their morphological parameters. This, in turn, offers valuable insights into their formation, evolutionary processes, and structural properties. We will further explore the impact of cooling flows on the growth of BCGs in the future.

\section{Conclusions}\label{sec:5}

As the most massive galaxies at the centers of clusters, BCGs play a crucial role in understanding galaxy formation and evolution, with their star-forming fraction serving as a key tracer of their growth and transformation. In this study, we utilize two large samples—221,256 BCGs with photometric redshifts and 55,765 BCGs with spectroscopic redshifts, spanning the range $0.1<z<0.8$—to investigate the evolution of the star-forming fraction of BCGs as a function of redshift, halo mass, and stellar mass. We further compare our results with field galaxies. Our goal is to explore the formation and evolutionary history of BCGs and to identify the key factors driving their mass growth. Our main findings are as follows:

\begin{enumerate}

\item Within the redshift range of 0.1–0.8, the $F_{\mathrm{sf}}$ for BCGs increases with redshift, showing a slow rise below $z \sim 0.4-0.5$ and more rapid increase above this range. Taking field galaxies as the comparison sample, the relative $F_{\mathrm{sf}}$ for BCGs displays a similar upward trend, with a more rapid increase observed at higher redshifts.

\item The $F_{\mathrm{sf}}$ of BCGs decreases with the increasing cluster halo mass and their stellar mass. At $z < 0.4$, the $F_{\mathrm{sf}}$ of BCGs remains low across all halo mass ranges, being nearly zero in high-mass halos, about 5\% in intermediate-mass halos, and around 10\% in low-mass halos. As redshift increases beyond 0.4, the $F_{\mathrm{sf}}$ rises sharply in all three halo mass bins, with the most rapid increase in low-mass halos. At $z > 0.4$, the average $F_{\mathrm{sf}}$ of BCGs in low-mass halos is 1.3 times that in intermediate-mass halos and 2.2 times that in high-mass halos, while in intermediate-mass halos it is 1.7 times that in high-mass halos.

\item Star-forming BCGs tend to exhibit larger projected offsets from the optical cluster density peak than quenching BCGs, indicating that large BCG–centre offsets trace recent halo mergers or ongoing assembly, whereas small offsets point to earlier assembly and more relaxed systems.

\item At the low stellar mass end, BCGs exhibit higher star-forming activity compared to field galaxies. We propose that this difference stems from variations in the mechanisms fueling star formation in BCGs across different redshifts. At low redshifts, the star-forming activity is likely predominantly driven by cooling flows, while at higher redshifts the star formation of BCGs is driven by mergers. Our sample indicates that the transition occurs around redshift 0.5.

\end{enumerate}

\section*{Acknowledgements}

The authors acknowledge the supports from National Key R\&D Program of China (grant Nos. 2023YFA1607804, 2022YFA1602902, 2023YFA1607800, 2023YFA1608100), the National Natural Science Foundation of China (NSFC; grant Nos. 12120101003, 12373010, 12173051, and 12233008) and China Manned Space Project (No. CMS-CSST-2025-A06). The authors also acknowledge the Strategic Priority Research Program of the Chinese Academy of Sciences with Grant Nos. XDB0550100 and XDB0550000.

\section*{Data Availability}

The photometric cluster catalogue used in this work is available from the ScienceDB repository at \href{https://doi.org/10.11922/sciencedb.o00069.00003}{doi:10.11922/sciencedb.o00069.00003}. The spectroscopic galaxy sample is publicly available from the DESI data release portal at \url{https://data.desi.lbl.gov/doc/releases/dr1/vac/stellar-mass-emline/}.




\bibliographystyle{mnras}
\bibliography{BCG_ref} 

@ARTICLE{zou2019,
       author = {{Zou}, Hu and {Gao}, Jinghua and {Zhou}, Xu and {Kong}, Xu},
        title = "{Photometric Redshifts and Stellar Masses for Galaxies from the DESI Legacy Imaging Surveys}",
      journal = {\apjs},
         year = 2019,
        month = may,
       volume = {242},
       number = {1},
          eid = {8},
        pages = {8},
          doi = {10.3847/1538-4365/ab1847},
       adsurl = {https://ui.adsabs.harvard.edu/abs/2019ApJS..242....8Z}
}

@ARTICLE{Overview_DESI_Dey.et.al.2019,
       author = {{Dey}, Arjun and {Schlegel}, David J. and {Lang}, Dustin and {Blum}, Robert and {Burleigh}, Kaylan and {Fan}, Xiaohui and {Findlay}, Joseph R. and {Finkbeiner}, Doug and {Herrera}, David and {Juneau}, St{\'e}phanie and {Landriau}, Martin and {Levi}, Michael and {McGreer}, Ian and {Meisner}, Aaron and {Myers}, Adam D. and {Moustakas}, John and {Nugent}, Peter and {Patej}, Anna and {Schlafly}, Edward F. and {Walker}, Alistair R. and {Valdes}, Francisco and {Weaver}, Benjamin A. and {Y{\`e}che}, Christophe and {Zou}, Hu and {Zhou}, Xu and {Abareshi}, Behzad and {Abbott}, T.~M.~C. and {Abolfathi}, Bela and {Aguilera}, C. and {Alam}, Shadab and {Allen}, Lori and {Alvarez}, A. and {Annis}, James and {Ansarinejad}, Behzad and {Aubert}, Marie and {Beechert}, Jacqueline and {Bell}, Eric F. and {BenZvi}, Segev Y. and {Beutler}, Florian and {Bielby}, Richard M. and {Bolton}, Adam S. and {Brice{\~n}o}, C{\'e}sar and {Buckley-Geer}, Elizabeth J. and {Butler}, Karen and {Calamida}, Annalisa and {Carlberg}, Raymond G. and {Carter}, Paul and {Casas}, Ricard and {Castander}, Francisco J. and {Choi}, Yumi and {Comparat}, Johan and {Cukanovaite}, Elena and {Delubac}, Timoth{\'e}e and {DeVries}, Kaitlin and {Dey}, Sharmila and {Dhungana}, Govinda and {Dickinson}, Mark and {Ding}, Zhejie and {Donaldson}, John B. and {Duan}, Yutong and {Duckworth}, Christopher J. and {Eftekharzadeh}, Sarah and {Eisenstein}, Daniel J. and {Etourneau}, Thomas and {Fagrelius}, Parker A. and {Farihi}, Jay and {Fitzpatrick}, Mike and {Font-Ribera}, Andreu and {Fulmer}, Leah and {G{\"a}nsicke}, Boris T. and {Gaztanaga}, Enrique and {George}, Koshy and {Gerdes}, David W. and {Gontcho}, Satya Gontcho A. and {Gorgoni}, Claudio and {Green}, Gregory and {Guy}, Julien and {Harmer}, Diane and {Hernandez}, M. and {Honscheid}, Klaus and {Huang}, Lijuan Wendy and {James}, David J. and {Jannuzi}, Buell T. and {Jiang}, Linhua and {Joyce}, Richard and {Karcher}, Armin and {Karkar}, Sonia and {Kehoe}, Robert and {Kneib}, Jean-Paul and {Kueter-Young}, Andrea and {Lan}, Ting-Wen and {Lauer}, Tod R. and {Le Guillou}, Laurent and {Le Van Suu}, Auguste and {Lee}, Jae Hyeon and {Lesser}, Michael and {Perreault Levasseur}, Laurence and {Li}, Ting S. and {Mann}, Justin L. and {Marshall}, Robert and {Mart{\'\i}nez-V{\'a}zquez}, C.~E. and {Martini}, Paul and {du Mas des Bourboux}, H{\'e}lion and {McManus}, Sean and {Meier}, Tobias Gabriel and {M{\'e}nard}, Brice and {Metcalfe}, Nigel and {Mu{\~n}oz-Guti{\'e}rrez}, Andrea and {Najita}, Joan and {Napier}, Kevin and {Narayan}, Gautham and {Newman}, Jeffrey A. and {Nie}, Jundan and {Nord}, Brian and {Norman}, Dara J. and {Olsen}, Knut A.~G. and {Paat}, Anthony and {Palanque-Delabrouille}, Nathalie and {Peng}, Xiyan and {Poppett}, Claire L. and {Poremba}, Megan R. and {Prakash}, Abhishek and {Rabinowitz}, David and {Raichoor}, Anand and {Rezaie}, Mehdi and {Robertson}, A.~N. and {Roe}, Natalie A. and {Ross}, Ashley J. and {Ross}, Nicholas P. and {Rudnick}, Gregory and {Safonova}, Sasha and {Saha}, Abhijit and {S{\'a}nchez}, F. Javier and {Savary}, Elodie and {Schweiker}, Heidi and {Scott}, Adam and {Seo}, Hee-Jong and {Shan}, Huanyuan and {Silva}, David R. and {Slepian}, Zachary and {Soto}, Christian and {Sprayberry}, David and {Staten}, Ryan and {Stillman}, Coley M. and {Stupak}, Robert J. and {Summers}, David L. and {Sien Tie}, Suk and {Tirado}, H. and {Vargas-Maga{\~n}a}, Mariana and {Vivas}, A. Katherina and {Wechsler}, Risa H. and {Williams}, Doug and {Yang}, Jinyi and {Yang}, Qian and {Yapici}, Tolga and {Zaritsky}, Dennis and {Zenteno}, A. and {Zhang}, Kai and {Zhang}, Tianmeng and {Zhou}, Rongpu and {Zhou}, Zhimin},
        title = "{Overview of the DESI Legacy Imaging Surveys}",
      journal = {\aj},
         year = 2019,
        month = may,
       volume = {157},
       number = {5},
          eid = {168},
        pages = {168},
          doi = {10.3847/1538-3881/ab089d},
archivePrefix = {arXiv},
       eprint = {1804.08657},
 primaryClass = {astro-ph.IM},
       adsurl = {https://ui.adsabs.harvard.edu/abs/2019AJ....157..168D}
}

@ARTICLE{BASS;Zou.et.al.2017,
       author = {{Zou}, Hu and {Zhou}, Xu and {Fan}, Xiaohui and {Zhang}, Tianmeng and {Zhou}, Zhimin and {Nie}, Jundan and {Peng}, Xiyan and {McGreer}, Ian and {Jiang}, Linhua and {Dey}, Arjun and {Fan}, Dongwei and {He}, Boliang and {Jiang}, Zhaoji and {Lang}, Dustin and {Lesser}, Michael and {Ma}, Jun and {Mao}, Shude and {Schlegel}, David and {Wang}, Jiali},
        title = "{Project Overview of the Beijing-Arizona Sky Survey}",
      journal = {\pasp},
         year = 2017,
        month = jun,
       volume = {129},
       number = {976},
        pages = {064101},
          doi = {10.1088/1538-3873/aa65ba},
archivePrefix = {arXiv},
       eprint = {1702.03653},
 primaryClass = {astro-ph.GA},
       adsurl = {https://ui.adsabs.harvard.edu/abs/2017PASP..129f4101Z}
}

@ARTICLE{zou2022_dr9,
       author = {{Zou}, Hu and {Sui}, Jipeng and {Xue}, Suijian and {Zhou}, Xu and {Ma}, Jun and {Zhou}, Zhimin and {Nie}, Jundan and {Zhang}, Tianmeng and {Feng}, Lu and {Shen}, Zhixia and {Wang}, Jiali},
        title = "{Photometric Redshifts and Galaxy Clusters for DES DR2, DESI DR9, and HSC-SSP PDR3 Data}",
      journal = {Research in Astronomy and Astrophysics},
         year = 2022,
        month = jun,
       volume = {22},
       number = {6},
          eid = {065001},
        pages = {065001},
          doi = {10.1088/1674-4527/ac6416},
archivePrefix = {arXiv},
       eprint = {2203.17035},
 primaryClass = {astro-ph.GA},
       adsurl = {https://ui.adsabs.harvard.edu/abs/2022RAA....22f5001Z}
}

@ARTICLE{beck_2016,
       author = {{Beck}, R{\'o}bert and {Dobos}, L{\'a}szl{\'o} and {Budav{\'a}ri}, Tam{\'a}s and {Szalay}, Alexander S. and {Csabai}, Istv{\'a}n},
        title = "{Photometric redshifts for the SDSS Data Release 12}",
      journal = {\mnras},
         year = 2016,
        month = aug,
       volume = {460},
       number = {2},
        pages = {1371-1381},
          doi = {10.1093/mnras/stw1009},
archivePrefix = {arXiv},
       eprint = {1603.09708},
 primaryClass = {astro-ph.GA},
       adsurl = {https://ui.adsabs.harvard.edu/abs/2016MNRAS.460.1371B}
}

@ARTICLE{Gao2018,
       author = {{Gao}, Jinghua and {Zou}, Hu and {Zhou}, Xu and {Kong}, Xu},
        title = "{A Photometric Redshift Catalog Based on SCUSS, SDSS, and WISE Surveys}",
      journal = {\apj},
         year = 2018,
        month = jul,
       volume = {862},
       number = {1},
          eid = {12},
        pages = {12},
          doi = {10.3847/1538-4357/aacbc6},
       adsurl = {https://ui.adsabs.harvard.edu/abs/2018ApJ...862...12G}
}

@ARTICLE{Gao2020cluster,
       author = {{Gao}, Jinghua and {Zou}, Hu and {Zhou}, Xu and {Kong}, Xu},
        title = "{A Catalog of Galaxy Clusters Identified from SCUSS, SDSS, and UNWISE}",
      journal = {\pasp},
         year = 2020,
        month = feb,
       volume = {132},
       number = {1008},
          eid = {024101},
        pages = {024101},
          doi = {10.1088/1538-3873/ab6151},
archivePrefix = {arXiv},
       eprint = {1912.10909},
 primaryClass = {astro-ph.GA},
       adsurl = {https://ui.adsabs.harvard.edu/abs/2020PASP..132b4101G}
}

@ARTICLE{zou2021cluster,
       author = {{Zou}, Hu and {Gao}, Jinghua and {Xu}, Xin and {Zhou}, Xu and {Ma}, Jun and {Zhou}, Zhimin and {Zhang}, Tianmeng and {Nie}, Jundan and {Wang}, Jiali and {Xue}, Suijian},
        title = "{Galaxy Clusters from the DESI Legacy Imaging Surveys. I. Cluster Detection}",
      journal = {\apjs},
         year = 2021,
        month = apr,
       volume = {253},
       number = {2},
          eid = {56},
        pages = {56},
          doi = {10.3847/1538-4365/abe5b0},
archivePrefix = {arXiv},
       eprint = {2101.12340},
 primaryClass = {astro-ph.GA},
       adsurl = {https://ui.adsabs.harvard.edu/abs/2021ApJS..253...56Z}
}

@ARTICLE{DESI_Collaboration_2016_a,
       author = {{DESI Collaboration} and {Aghamousa}, Amir and {Aguilar}, Jessica and {Ahlen}, Steve and {Alam}, Shadab and {Allen}, Lori E. and {Allende Prieto}, Carlos and {Annis}, James and {Bailey}, Stephen and {Balland}, Christophe and {Ballester}, Otger and {Baltay}, Charles and {Beaufore}, Lucas and {Bebek}, Chris and {Beers}, Timothy C. and {Bell}, Eric F. and {Bernal}, Jos{\'e} Luis and {Besuner}, Robert and {Beutler}, Florian and {Blake}, Chris and {Bleuler}, Hannes and {Blomqvist}, Michael and {Blum}, Robert and {Bolton}, Adam S. and {Briceno}, Cesar and {Brooks}, David and {Brownstein}, Joel R. and {Buckley-Geer}, Elizabeth and {Burden}, Angela and {Burtin}, Etienne and {Busca}, Nicolas G. and {Cahn}, Robert N. and {Cai}, Yan-Chuan and {Cardiel-Sas}, Laia and {Carlberg}, Raymond G. and {Carton}, Pierre-Henri and {Casas}, Ricard and {Castander}, Francisco J. and {Cervantes-Cota}, Jorge L. and {Claybaugh}, Todd M. and {Close}, Madeline and {Coker}, Carl T. and {Cole}, Shaun and {Comparat}, Johan and {Cooper}, Andrew P. and {Cousinou}, M. -C. and {Crocce}, Martin and {Cuby}, Jean-Gabriel and {Cunningham}, Daniel P. and {Davis}, Tamara M. and {Dawson}, Kyle S. and {de la Macorra}, Axel and {De Vicente}, Juan and {Delubac}, Timoth{\'e}e and {Derwent}, Mark and {Dey}, Arjun and {Dhungana}, Govinda and {Ding}, Zhejie and {Doel}, Peter and {Duan}, Yutong T. and {Ealet}, Anne and {Edelstein}, Jerry and {Eftekharzadeh}, Sarah and {Eisenstein}, Daniel J. and {Elliott}, Ann and {Escoffier}, St{\'e}phanie and {Evatt}, Matthew and {Fagrelius}, Parker and {Fan}, Xiaohui and {Fanning}, Kevin and {Farahi}, Arya and {Farihi}, Jay and {Favole}, Ginevra and {Feng}, Yu and {Fernandez}, Enrique and {Findlay}, Joseph R. and {Finkbeiner}, Douglas P. and {Fitzpatrick}, Michael J. and {Flaugher}, Brenna and {Flender}, Samuel and {Font-Ribera}, Andreu and {Forero-Romero}, Jaime E. and {Fosalba}, Pablo and {Frenk}, Carlos S. and {Fumagalli}, Michele and {Gaensicke}, Boris T. and {Gallo}, Giuseppe and {Garcia-Bellido}, Juan and {Gaztanaga}, Enrique and {Pietro Gentile Fusillo}, Nicola and {Gerard}, Terry and {Gershkovich}, Irena and {Giannantonio}, Tommaso and {Gillet}, Denis and {Gonzalez-de-Rivera}, Guillermo and {Gonzalez-Perez}, Violeta and {Gott}, Shelby and {Graur}, Or and {Gutierrez}, Gaston and {Guy}, Julien and {Habib}, Salman and {Heetderks}, Henry and {Heetderks}, Ian and {Heitmann}, Katrin and {Hellwing}, Wojciech A. and {Herrera}, David A. and {Ho}, Shirley and {Holland}, Stephen and {Honscheid}, Klaus and {Huff}, Eric and {Hutchinson}, Timothy A. and {Huterer}, Dragan and {Hwang}, Ho Seong and {Illa Laguna}, Joseph Maria and {Ishikawa}, Yuzo and {Jacobs}, Dianna and {Jeffrey}, Niall and {Jelinsky}, Patrick and {Jennings}, Elise and {Jiang}, Linhua and {Jimenez}, Jorge and {Johnson}, Jennifer and {Joyce}, Richard and {Jullo}, Eric and {Juneau}, St{\'e}phanie and {Kama}, Sami and {Karcher}, Armin and {Karkar}, Sonia and {Kehoe}, Robert and {Kennamer}, Noble and {Kent}, Stephen and {Kilbinger}, Martin and {Kim}, Alex G. and {Kirkby}, David and {Kisner}, Theodore and {Kitanidis}, Ellie and {Kneib}, Jean-Paul and {Koposov}, Sergey and {Kovacs}, Eve and {Koyama}, Kazuya and {Kremin}, Anthony and {Kron}, Richard and {Kronig}, Luzius and {Kueter-Young}, Andrea and {Lacey}, Cedric G. and {Lafever}, Robin and {Lahav}, Ofer and {Lambert}, Andrew and {Lampton}, Michael and {Landriau}, Martin and {Lang}, Dustin and {Lauer}, Tod R. and {Le Goff}, Jean-Marc and {Le Guillou}, Laurent and {Le Van Suu}, Auguste and {Lee}, Jae Hyeon and {Lee}, Su-Jeong and {Leitner}, Daniela and {Lesser}, Michael and {Levi}, Michael E. and {L'Huillier}, Benjamin and {Li}, Baojiu and {Liang}, Ming and {Lin}, Huan and {Linder}, Eric and {Loebman}, Sarah R. and {Luki{\'c}}, Zarija and {Ma}, Jun and {MacCrann}, Niall and {Magneville}, Christophe and {Makarem}, Laleh and {Manera}, Marc and {Manser}, Christopher J. and {Marshall}, Robert and {Martini}, Paul and {Massey}, Richard and {Matheson}, Thomas and {McCauley}, Jeremy and {McDonald}, Patrick and {McGreer}, Ian D. and {Meisner}, Aaron and {Metcalfe}, Nigel and {Miller}, Timothy N. and {Miquel}, Ramon and {Moustakas}, John and {Myers}, Adam and {Naik}, Milind and {Newman}, Jeffrey A. and {Nichol}, Robert C. and {Nicola}, Andrina and {Nicolati da Costa}, Luiz and {Nie}, Jundan and {Niz}, Gustavo and {Norberg}, Peder and {Nord}, Brian and {Norman}, Dara and {Nugent}, Peter and {O'Brien}, Thomas and {Oh}, Minji and {Olsen}, Knut A.~G. and {Padilla}, Cristobal and {Padmanabhan}, Hamsa and {Padmanabhan}, Nikhil and {Palanque-Delabrouille}, Nathalie and {Palmese}, Antonella and {Pappalardo}, Daniel and {P{\^a}ris}, Isabelle and {Park}, Changbom and {Patej}, Anna and {Peacock}, John A. and {Peiris}, Hiranya V. and {Peng}, Xiyan and {Percival}, Will J. and {Perruchot}, Sandrine and {Pieri}, Matthew M. and {Pogge}, Richard and {Pollack}, Jennifer E. and {Poppett}, Claire and {Prada}, Francisco and {Prakash}, Abhishek and {Probst}, Ronald G. and {Rabinowitz}, David and {Raichoor}, Anand and {Ree}, Chang Hee and {Refregier}, Alexandre and {Regal}, Xavier and {Reid}, Beth and {Reil}, Kevin and {Rezaie}, Mehdi and {Rockosi}, Constance M. and {Roe}, Natalie and {Ronayette}, Samuel and {Roodman}, Aaron and {Ross}, Ashley J. and {Ross}, Nicholas P. and {Rossi}, Graziano and {Rozo}, Eduardo and {Ruhlmann-Kleider}, Vanina and {Rykoff}, Eli S. and {Sabiu}, Cristiano and {Samushia}, Lado and {Sanchez}, Eusebio and {Sanchez}, Javier and {Schlegel}, David J. and {Schneider}, Michael and {Schubnell}, Michael and {Secroun}, Aur{\'e}lia and {Seljak}, Uros and {Seo}, Hee-Jong and {Serrano}, Santiago and {Shafieloo}, Arman and {Shan}, Huanyuan and {Sharples}, Ray and {Sholl}, Michael J. and {Shourt}, William V. and {Silber}, Joseph H. and {Silva}, David R. and {Sirk}, Martin M. and {Slosar}, Anze and {Smith}, Alex and {Smoot}, George F. and {Som}, Debopam and {Song}, Yong-Seon and {Sprayberry}, David and {Staten}, Ryan and {Stefanik}, Andy and {Tarle}, Gregory and {Sien Tie}, Suk and {Tinker}, Jeremy L. and {Tojeiro}, Rita and {Valdes}, Francisco and {Valenzuela}, Octavio and {Valluri}, Monica and {Vargas-Magana}, Mariana and {Verde}, Licia and {Walker}, Alistair R. and {Wang}, Jiali and {Wang}, Yuting and {Weaver}, Benjamin A. and {Weaverdyck}, Curtis and {Wechsler}, Risa H. and {Weinberg}, David H. and {White}, Martin and {Yang}, Qian and {Yeche}, Christophe and {Zhang}, Tianmeng and {Zhao}, Gong-Bo and {Zheng}, Yi and {Zhou}, Xu and {Zhou}, Zhimin and {Zhu}, Yaling and {Zou}, Hu and {Zu}, Ying},
        title = "{The DESI Experiment Part I: Science,Targeting, and Survey Design}",
      journal = {arXiv e-prints},
         year = 2016,
        month = oct,
          eid = {arXiv:1611.00036},
        pages = {arXiv:1611.00036},
          doi = {10.48550/arXiv.1611.00036},
archivePrefix = {arXiv},
       eprint = {1611.00036},
 primaryClass = {astro-ph.IM},
       adsurl = {https://ui.adsabs.harvard.edu/abs/2016arXiv161100036D}
}

@ARTICLE{DESI_Collaboration_2016_b,
       author = {{DESI Collaboration} and {Aghamousa}, Amir and {Aguilar}, Jessica and {Ahlen}, Steve and {Alam}, Shadab and {Allen}, Lori E. and {Allende Prieto}, Carlos and {Annis}, James and {Bailey}, Stephen and {Balland}, Christophe and {Ballester}, Otger and {Baltay}, Charles and {Beaufore}, Lucas and {Bebek}, Chris and {Beers}, Timothy C. and {Bell}, Eric F. and {Bernal}, Jos{\'e} Luis and {Besuner}, Robert and {Beutler}, Florian and {Blake}, Chris and {Bleuler}, Hannes and {Blomqvist}, Michael and {Blum}, Robert and {Bolton}, Adam S. and {Briceno}, Cesar and {Brooks}, David and {Brownstein}, Joel R. and {Buckley-Geer}, Elizabeth and {Burden}, Angela and {Burtin}, Etienne and {Busca}, Nicolas G. and {Cahn}, Robert N. and {Cai}, Yan-Chuan and {Cardiel-Sas}, Laia and {Carlberg}, Raymond G. and {Carton}, Pierre-Henri and {Casas}, Ricard and {Castander}, Francisco J. and {Cervantes-Cota}, Jorge L. and {Claybaugh}, Todd M. and {Close}, Madeline and {Coker}, Carl T. and {Cole}, Shaun and {Comparat}, Johan and {Cooper}, Andrew P. and {Cousinou}, M. -C. and {Crocce}, Martin and {Cuby}, Jean-Gabriel and {Cunningham}, Daniel P. and {Davis}, Tamara M. and {Dawson}, Kyle S. and {de la Macorra}, Axel and {De Vicente}, Juan and {Delubac}, Timoth{\'e}e and {Derwent}, Mark and {Dey}, Arjun and {Dhungana}, Govinda and {Ding}, Zhejie and {Doel}, Peter and {Duan}, Yutong T. and {Ealet}, Anne and {Edelstein}, Jerry and {Eftekharzadeh}, Sarah and {Eisenstein}, Daniel J. and {Elliott}, Ann and {Escoffier}, St{\'e}phanie and {Evatt}, Matthew and {Fagrelius}, Parker and {Fan}, Xiaohui and {Fanning}, Kevin and {Farahi}, Arya and {Farihi}, Jay and {Favole}, Ginevra and {Feng}, Yu and {Fernandez}, Enrique and {Findlay}, Joseph R. and {Finkbeiner}, Douglas P. and {Fitzpatrick}, Michael J. and {Flaugher}, Brenna and {Flender}, Samuel and {Font-Ribera}, Andreu and {Forero-Romero}, Jaime E. and {Fosalba}, Pablo and {Frenk}, Carlos S. and {Fumagalli}, Michele and {Gaensicke}, Boris T. and {Gallo}, Giuseppe and {Garcia-Bellido}, Juan and {Gaztanaga}, Enrique and {Pietro Gentile Fusillo}, Nicola and {Gerard}, Terry and {Gershkovich}, Irena and {Giannantonio}, Tommaso and {Gillet}, Denis and {Gonzalez-de-Rivera}, Guillermo and {Gonzalez-Perez}, Violeta and {Gott}, Shelby and {Graur}, Or and {Gutierrez}, Gaston and {Guy}, Julien and {Habib}, Salman and {Heetderks}, Henry and {Heetderks}, Ian and {Heitmann}, Katrin and {Hellwing}, Wojciech A. and {Herrera}, David A. and {Ho}, Shirley and {Holland}, Stephen and {Honscheid}, Klaus and {Huff}, Eric and {Hutchinson}, Timothy A. and {Huterer}, Dragan and {Hwang}, Ho Seong and {Illa Laguna}, Joseph Maria and {Ishikawa}, Yuzo and {Jacobs}, Dianna and {Jeffrey}, Niall and {Jelinsky}, Patrick and {Jennings}, Elise and {Jiang}, Linhua and {Jimenez}, Jorge and {Johnson}, Jennifer and {Joyce}, Richard and {Jullo}, Eric and {Juneau}, St{\'e}phanie and {Kama}, Sami and {Karcher}, Armin and {Karkar}, Sonia and {Kehoe}, Robert and {Kennamer}, Noble and {Kent}, Stephen and {Kilbinger}, Martin and {Kim}, Alex G. and {Kirkby}, David and {Kisner}, Theodore and {Kitanidis}, Ellie and {Kneib}, Jean-Paul and {Koposov}, Sergey and {Kovacs}, Eve and {Koyama}, Kazuya and {Kremin}, Anthony and {Kron}, Richard and {Kronig}, Luzius and {Kueter-Young}, Andrea and {Lacey}, Cedric G. and {Lafever}, Robin and {Lahav}, Ofer and {Lambert}, Andrew and {Lampton}, Michael and {Landriau}, Martin and {Lang}, Dustin and {Lauer}, Tod R. and {Le Goff}, Jean-Marc and {Le Guillou}, Laurent and {Le Van Suu}, Auguste and {Lee}, Jae Hyeon and {Lee}, Su-Jeong and {Leitner}, Daniela and {Lesser}, Michael and {Levi}, Michael E. and {L'Huillier}, Benjamin and {Li}, Baojiu and {Liang}, Ming and {Lin}, Huan and {Linder}, Eric and {Loebman}, Sarah R. and {Luki{\'c}}, Zarija and {Ma}, Jun and {MacCrann}, Niall and {Magneville}, Christophe and {Makarem}, Laleh and {Manera}, Marc and {Manser}, Christopher J. and {Marshall}, Robert and {Martini}, Paul and {Massey}, Richard and {Matheson}, Thomas and {McCauley}, Jeremy and {McDonald}, Patrick and {McGreer}, Ian D. and {Meisner}, Aaron and {Metcalfe}, Nigel and {Miller}, Timothy N. and {Miquel}, Ramon and {Moustakas}, John and {Myers}, Adam and {Naik}, Milind and {Newman}, Jeffrey A. and {Nichol}, Robert C. and {Nicola}, Andrina and {Nicolati da Costa}, Luiz and {Nie}, Jundan and {Niz}, Gustavo and {Norberg}, Peder and {Nord}, Brian and {Norman}, Dara and {Nugent}, Peter and {O'Brien}, Thomas and {Oh}, Minji and {Olsen}, Knut A.~G. and {Padilla}, Cristobal and {Padmanabhan}, Hamsa and {Padmanabhan}, Nikhil and {Palanque-Delabrouille}, Nathalie and {Palmese}, Antonella and {Pappalardo}, Daniel and {P{\^a}ris}, Isabelle and {Park}, Changbom and {Patej}, Anna and {Peacock}, John A. and {Peiris}, Hiranya V. and {Peng}, Xiyan and {Percival}, Will J. and {Perruchot}, Sandrine and {Pieri}, Matthew M. and {Pogge}, Richard and {Pollack}, Jennifer E. and {Poppett}, Claire and {Prada}, Francisco and {Prakash}, Abhishek and {Probst}, Ronald G. and {Rabinowitz}, David and {Raichoor}, Anand and {Ree}, Chang Hee and {Refregier}, Alexandre and {Regal}, Xavier and {Reid}, Beth and {Reil}, Kevin and {Rezaie}, Mehdi and {Rockosi}, Constance M. and {Roe}, Natalie and {Ronayette}, Samuel and {Roodman}, Aaron and {Ross}, Ashley J. and {Ross}, Nicholas P. and {Rossi}, Graziano and {Rozo}, Eduardo and {Ruhlmann-Kleider}, Vanina and {Rykoff}, Eli S. and {Sabiu}, Cristiano and {Samushia}, Lado and {Sanchez}, Eusebio and {Sanchez}, Javier and {Schlegel}, David J. and {Schneider}, Michael and {Schubnell}, Michael and {Secroun}, Aur{\'e}lia and {Seljak}, Uros and {Seo}, Hee-Jong and {Serrano}, Santiago and {Shafieloo}, Arman and {Shan}, Huanyuan and {Sharples}, Ray and {Sholl}, Michael J. and {Shourt}, William V. and {Silber}, Joseph H. and {Silva}, David R. and {Sirk}, Martin M. and {Slosar}, Anze and {Smith}, Alex and {Smoot}, George F. and {Som}, Debopam and {Song}, Yong-Seon and {Sprayberry}, David and {Staten}, Ryan and {Stefanik}, Andy and {Tarle}, Gregory and {Sien Tie}, Suk and {Tinker}, Jeremy L. and {Tojeiro}, Rita and {Valdes}, Francisco and {Valenzuela}, Octavio and {Valluri}, Monica and {Vargas-Magana}, Mariana and {Verde}, Licia and {Walker}, Alistair R. and {Wang}, Jiali and {Wang}, Yuting and {Weaver}, Benjamin A. and {Weaverdyck}, Curtis and {Wechsler}, Risa H. and {Weinberg}, David H. and {White}, Martin and {Yang}, Qian and {Yeche}, Christophe and {Zhang}, Tianmeng and {Zhao}, Gong-Bo and {Zheng}, Yi and {Zhou}, Xu and {Zhou}, Zhimin and {Zhu}, Yaling and {Zou}, Hu and {Zu}, Ying},
        title = "{The DESI Experiment Part II: Instrument Design}",
      journal = {arXiv e-prints},
         year = 2016,
        month = oct,
          eid = {arXiv:1611.00037},
        pages = {arXiv:1611.00037},
          doi = {10.48550/arXiv.1611.00037},
archivePrefix = {arXiv},
       eprint = {1611.00037},
 primaryClass = {astro-ph.IM},
       adsurl = {https://ui.adsabs.harvard.edu/abs/2016arXiv161100037D}
}

@ARTICLE{DESI_Collaboration_2024_a,
       author = {{DESI Collaboration} and {Adame}, A.~G. and {Aguilar}, J. and {Ahlen}, S. and {Alam}, S. and {Aldering}, G. and {Alexander}, D.~M. and {Alfarsy}, R. and {Allende Prieto}, C. and {Alvarez}, M. and {Alves}, O. and {Anand}, A. and {Andrade-Oliveira}, F. and {Armengaud}, E. and {Asorey}, J. and {Avila}, S. and {Aviles}, A. and {Bailey}, S. and {Balaguera-Antol{\'\i}nez}, A. and {Ballester}, O. and {Baltay}, C. and {Bault}, A. and {Bautista}, J. and {Behera}, J. and {Beltran}, S.~F. and {BenZvi}, S. and {Beraldo e Silva}, L. and {Bermejo-Climent}, J.~R. and {Berti}, A. and {Besuner}, R. and {Beutler}, F. and {Bianchi}, D. and {Blake}, C. and {Blum}, R. and {Bolton}, A.~S. and {Brieden}, S. and {Brodzeller}, A. and {Brooks}, D. and {Brown}, Z. and {Buckley-Geer}, E. and {Burtin}, E. and {Cabayol-Garcia}, L. and {Cai}, Z. and {Canning}, R. and {Cardiel-Sas}, L. and {Carnero Rosell}, A. and {Castander}, F.~J. and {Cervantes-Cota}, J.~L. and {Chabanier}, S. and {Chaussidon}, E. and {Chaves-Montero}, J. and {Chen}, S. and {Chen}, X. and {Chuang}, C. and {Claybaugh}, T. and {Cole}, S. and {Cooper}, A.~P. and {Cuceu}, A. and {Davis}, T.~M. and {Dawson}, K. and {de Belsunce}, R. and {de la Cruz}, R. and {de la Macorra}, A. and {de Mattia}, A. and {Demina}, R. and {Demirbozan}, U. and {DeRose}, J. and {Dey}, A. and {Dey}, B. and {Dhungana}, G. and {Ding}, J. and {Ding}, Z. and {Doel}, P. and {Doshi}, R. and {Douglass}, K. and {Edge}, A. and {Eftekharzadeh}, S. and {Eisenstein}, D.~J. and {Elliott}, A. and {Escoffier}, S. and {Fagrelius}, P. and {Fan}, X. and {Fanning}, K. and {Fawcett}, V.~A. and {Ferraro}, S. and {Ereza}, J. and {Flaugher}, B. and {Font-Ribera}, A. and {Forero-S{\'a}nchez}, D. and {Forero-Romero}, J.~E. and {Frenk}, C.~S. and {G{\"a}nsicke}, B.~T. and {Garc{\'\i}a}, L. {\'A}. and {Garc{\'\i}a-Bellido}, J. and {Garcia-Quintero}, C. and {Garrison}, L.~H. and {Gil-Mar{\'\i}n}, H. and {Golden-Marx}, J. and {Gontcho A Gontcho}, S. and {Gonzalez-Morales}, A.~X. and {Gonzalez-Perez}, V. and {Gordon}, C. and {Graur}, O. and {Green}, D. and {Gruen}, D. and {Guy}, J. and {Hadzhiyska}, B. and {Hahn}, C. and {Han}, J.~J. and {Hanif}, M.~M.~S. and {Herrera-Alcantar}, H.~K. and {Honscheid}, K. and {Hou}, J. and {Howlett}, C. and {Huterer}, D. and {Ir{\v{s}}i{\v{c}}}, V. and {Ishak}, M. and {Jana}, A. and {Jiang}, L. and {Jimenez}, J. and {Jing}, Y.~P. and {Joudaki}, S. and {Jullo}, E. and {Joyce}, R. and {Juneau}, S. and {Kizhuprakkat}, N. and {Kara{\c{c}}ayl{\i}}, N.~G. and {Karim}, T. and {Kehoe}, R. and {Kent}, S. and {Khederlarian}, A. and {Kim}, S. and {Kirkby}, D. and {Kisner}, T. and {Kitaura}, F. and {Kneib}, J. and {Koposov}, S.~E. and {Kov{\'a}cs}, A. and {Kremin}, A. and {Krolewski}, A. and {L'Huillier}, B. and {Lahav}, O. and {Lambert}, A. and {Lamman}, C. and {Lan}, T. -W. and {Landriau}, M. and {Lang}, D. and {Lange}, J.~U. and {Lasker}, J. and {Le Guillou}, L. and {Leauthaud}, A. and {Levi}, M.~E. and {Li}, T.~S. and {Linder}, E. and {Lyons}, A. and {Magneville}, C. and {Manera}, M. and {Manser}, C.~J. and {Margala}, D. and {Martini}, P. and {McDonald}, P. and {Medina}, G.~E. and {Medina-Varela}, L. and {Meisner}, A. and {Mena-Fern{\'a}ndez}, J. and {Meneses-Rizo}, J. and {Mezcua}, M. and {Miquel}, R. and {Montero-Camacho}, P. and {Moon}, J. and {Moore}, S. and {Moustakas}, J. and {Mueller}, E. and {Mundet}, J. and {Mu{\~n}oz-Guti{\'e}rrez}, A. and {Myers}, A.~D. and {Nadathur}, S. and {Napolitano}, L. and {Neveux}, R. and {Newman}, J.~A. and {Nie}, J. and {Niz}, G. and {Norberg}, P. and {Noriega}, H.~E. and {Paillas}, E. and {Palanque-Delabrouille}, N. and {Palmese}, A. and {Zhiwei}, P. and {Parkinson}, D. and {Penmetsa}, S. and {Percival}, W.~J. and {P{\'e}rez-Fern{\'a}ndez}, A. and {P{\'e}rez-R{\`a}fols}, I. and {Pieri}, M. and {Poppett}, C. and {Porredon}, A. and {Prada}, F. and {Pucha}, R. and {Raichoor}, A. and {Ram{\'\i}rez-P{\'e}rez}, C. and {Ramirez-Solano}, S. and {Rashkovetskyi}, M. and {Ravoux}, C. and {Rocher}, A. and {Rockosi}, C. and {Ross}, A.~J. and {Rossi}, G. and {Ruggeri}, R. and {Ruhlmann-Kleider}, V. and {Sabiu}, C.~G. and {Said}, K. and {Saintonge}, A. and {Samushia}, L. and {Sanchez}, E. and {Saulder}, C. and {Schaan}, E. and {Schlafly}, E.~F. and {Schlegel}, D. and {Scholte}, D. and {Schubnell}, M. and {Seo}, H. and {Shafieloo}, A. and {Sharples}, R. and {Sheu}, W. and {Silber}, J. and {Sinigaglia}, F. and {Siudek}, M. and {Slepian}, Z. and {Smith}, A. and {Sprayberry}, D. and {Stephey}, L. and {Su{\'a}rez-P{\'e}rez}, J. and {Sun}, Z. and {Tan}, T. and {Tarl{\'e}}, G. and {Tojeiro}, R. and {Ure{\~n}a-L{\'o}pez}, L.~A. and {Vaisakh}, R. and {Valcin}, D. and {Valdes}, F. and {Valluri}, M. and {Vargas-Maga{\~n}a}, M. and {Variu}, A. and {Verde}, L. and {Walther}, M. and {Wang}, B. and {Wang}, M.~S. and {Weaver}, B.~A. and {Weaverdyck}, N. and {Wechsler}, R.~H. and {White}, M. and {Xie}, Y. and {Yang}, J. and {Y{\`e}che}, C. and {Yu}, J. and {Yuan}, S. and {Zhang}, H. and {Zhang}, Z. and {Zhao}, C. and {Zheng}, Z. and {Zhou}, R. and {Zhou}, Z. and {Zou}, H. and {Zou}, S. and {Zu}, Y. and {DESI Collaboration}},
        title = "{Validation of the Scientific Program for the Dark Energy Spectroscopic Instrument}",
      journal = {\aj},
         year = 2024,
        month = feb,
       volume = {167},
       number = {2},
          eid = {62},
        pages = {62},
          doi = {10.3847/1538-3881/ad0b08},
archivePrefix = {arXiv},
       eprint = {2306.06307},
 primaryClass = {astro-ph.CO},
       adsurl = {https://ui.adsabs.harvard.edu/abs/2024AJ....167...62D}
}

@ARTICLE{2DESI_Collaboration_2024_b,
       author = {{DESI Collaboration} and {Adame}, A.~G. and {Aguilar}, J. and {Ahlen}, S. and {Alam}, S. and {Aldering}, G. and {Alexander}, D.~M. and {Alfarsy}, R. and {Allende Prieto}, C. and {Alvarez}, M. and {Alves}, O. and {Anand}, A. and {Andrade-Oliveira}, F. and {Armengaud}, E. and {Asorey}, J. and {Avila}, S. and {Aviles}, A. and {Bailey}, S. and {Balaguera-Antol{\'\i}nez}, A. and {Ballester}, O. and {Baltay}, C. and {Bault}, A. and {Bautista}, J. and {Behera}, J. and {Beltran}, S.~F. and {BenZvi}, S. and {Beraldo e Silva}, L. and {Bermejo-Climent}, J.~R. and {Berti}, A. and {Besuner}, R. and {Beutler}, F. and {Bianchi}, D. and {Blake}, C. and {Blum}, R. and {Bolton}, A.~S. and {Brieden}, S. and {Brodzeller}, A. and {Brooks}, D. and {Brown}, Z. and {Buckley-Geer}, E. and {Burtin}, E. and {Cabayol-Garcia}, L. and {Cai}, Z. and {Canning}, R. and {Cardiel-Sas}, L. and {Carnero Rosell}, A. and {Castander}, F.~J. and {Cervantes-Cota}, J.~L. and {Chabanier}, S. and {Chaussidon}, E. and {Chaves-Montero}, J. and {Chen}, S. and {Chen}, X. and {Chuang}, C. and {Claybaugh}, T. and {Cole}, S. and {Cooper}, A.~P. and {Cuceu}, A. and {Davis}, T.~M. and {Dawson}, K. and {de Belsunce}, R. and {de la Cruz}, R. and {de la Macorra}, A. and {Della Costa}, J. and {de Mattia}, A. and {Demina}, R. and {Demirbozan}, U. and {DeRose}, J. and {Dey}, A. and {Dey}, B. and {Dhungana}, G. and {Ding}, J. and {Ding}, Z. and {Doel}, P. and {Doshi}, R. and {Douglass}, K. and {Edge}, A. and {Eftekharzadeh}, S. and {Eisenstein}, D.~J. and {Elliott}, A. and {Ereza}, J. and {Escoffier}, S. and {Fagrelius}, P. and {Fan}, X. and {Fanning}, K. and {Fawcett}, V.~A. and {Ferraro}, S. and {Flaugher}, B. and {Font-Ribera}, A. and {Forero-Romero}, J.~E. and {Forero-S{\'a}nchez}, D. and {Frenk}, C.~S. and {G{\"a}nsicke}, B.~T. and {Garc{\'\i}a}, L. {\'A}. and {Garc{\'\i}a-Bellido}, J. and {Garcia-Quintero}, C. and {Garrison}, L.~H. and {Gil-Mar{\'\i}n}, H. and {Golden-Marx}, J. and {Gontcho A Gontcho}, S. and {Gonzalez-Morales}, A.~X. and {Gonzalez-Perez}, V. and {Gordon}, C. and {Graur}, O. and {Green}, D. and {Gruen}, D. and {Guy}, J. and {Hadzhiyska}, B. and {Hahn}, C. and {Han}, J.~J. and {Hanif}, M.~M.~S. and {Herrera-Alcantar}, H.~K. and {Honscheid}, K. and {Hou}, J. and {Howlett}, C. and {Huterer}, D. and {Ir{\v{s}}i{\v{c}}}, V. and {Ishak}, M. and {Jacques}, A. and {Jana}, A. and {Jiang}, L. and {Jimenez}, J. and {Jing}, Y.~P. and {Joudaki}, S. and {Joyce}, R. and {Jullo}, E. and {Juneau}, S. and {Kara{\c{c}}ayl{\i}}, N.~G. and {Karim}, T. and {Kehoe}, R. and {Kent}, S. and {Khederlarian}, A. and {Kim}, S. and {Kirkby}, D. and {Kisner}, T. and {Kitaura}, F. and {Kizhuprakkat}, N. and {Kneib}, J. and {Koposov}, S.~E. and {Kov{\'a}cs}, A. and {Kremin}, A. and {Krolewski}, A. and {L'Huillier}, B. and {Lahav}, O. and {Lambert}, A. and {Lamman}, C. and {Lan}, T. -W. and {Landriau}, M. and {Lang}, D. and {Lange}, J.~U. and {Lasker}, J. and {Leauthaud}, A. and {Le Guillou}, L. and {Levi}, M.~E. and {Li}, T.~S. and {Linder}, E. and {Lyons}, A. and {Magneville}, C. and {Manera}, M. and {Manser}, C.~J. and {Margala}, D. and {Martini}, P. and {McDonald}, P. and {Medina}, G.~E. and {Medina-Varela}, L. and {Meisner}, A. and {Mena-Fern{\'a}ndez}, J. and {Meneses-Rizo}, J. and {Mezcua}, M. and {Miquel}, R. and {Montero-Camacho}, P. and {Moon}, J. and {Moore}, S. and {Moustakas}, J. and {Mueller}, E. and {Mundet}, J. and {Mu{\~n}oz-Guti{\'e}rrez}, A. and {Myers}, A.~D. and {Nadathur}, S. and {Napolitano}, L. and {Neveux}, R. and {Newman}, J.~A. and {Nie}, J. and {Nikutta}, R. and {Niz}, G. and {Norberg}, P. and {Noriega}, H.~E. and {Paillas}, E. and {Palanque-Delabrouille}, N. and {Palmese}, A. and {Pan}, Z. and {Parkinson}, D. and {Penmetsa}, S. and {Percival}, W.~J. and {P{\'e}rez-Fern{\'a}ndez}, A. and {P{\'e}rez-R{\`a}fols}, I. and {Pieri}, M. and {Poppett}, C. and {Porredon}, A. and {Pothier}, S. and {Prada}, F. and {Pucha}, R. and {Raichoor}, A. and {Ram{\'\i}rez-P{\'e}rez}, C. and {Ramirez-Solano}, S. and {Rashkovetskyi}, M. and {Ravoux}, C. and {Rocher}, A. and {Rockosi}, C. and {Ross}, A.~J. and {Rossi}, G. and {Ruggeri}, R. and {Ruhlmann-Kleider}, V. and {Sabiu}, C.~G. and {Said}, K. and {Saintonge}, A. and {Samushia}, L. and {Sanchez}, E. and {Saulder}, C. and {Schaan}, E. and {Schlafly}, E.~F. and {Schlegel}, D. and {Scholte}, D. and {Schubnell}, M. and {Seo}, H. and {Shafieloo}, A. and {Sharples}, R. and {Sheu}, W. and {Silber}, J. and {Sinigaglia}, F. and {Siudek}, M. and {Slepian}, Z. and {Smith}, A. and {Soumagnac}, M.~T. and {Sprayberry}, D. and {Stephey}, L. and {Su{\'a}rez-P{\'e}rez}, J. and {Sun}, Z. and {Tan}, T. and {Tarl{\'e}}, G. and {Tojeiro}, R. and {Ure{\~n}a-L{\'o}pez}, L.~A. and {Vaisakh}, R. and {Valcin}, D. and {Valdes}, F. and {Valluri}, M. and {Vargas-Maga{\~n}a}, M. and {Variu}, A. and {Verde}, L. and {Walther}, M. and {Wang}, B. and {Wang}, M.~S. and {Weaver}, B.~A. and {Weaverdyck}, N. and {Wechsler}, R.~H. and {White}, M. and {Xie}, Y. and {Yang}, J. and {Y{\`e}che}, C. and {Yu}, J. and {Yuan}, S. and {Zhang}, H. and {Zhang}, Z. and {Zhao}, C. and {Zheng}, Z. and {Zhou}, R. and {Zhou}, Z. and {Zou}, H. and {Zou}, S. and {Zu}, Y.},
        title = "{The Early Data Release of the Dark Energy Spectroscopic Instrument}",
      journal = {\aj},
         year = 2024,
        month = aug,
       volume = {168},
       number = {2},
          eid = {58},
        pages = {58},
          doi = {10.3847/1538-3881/ad3217},
archivePrefix = {arXiv},
       eprint = {2306.06308},
 primaryClass = {astro-ph.CO},
       adsurl = {https://ui.adsabs.harvard.edu/abs/2024AJ....168...58D}
}

@ARTICLE{DESI_Collaboration_2022,
       author = {{DESI Collaboration} and {Abareshi}, B. and {Aguilar}, J. and {Ahlen}, S. and {Alam}, Shadab and {Alexander}, David M. and {Alfarsy}, R. and {Allen}, L. and {Allende Prieto}, C. and {Alves}, O. and {Ameel}, J. and {Armengaud}, E. and {Asorey}, J. and {Aviles}, Alejandro and {Bailey}, S. and {Balaguera-Antol{\'\i}nez}, A. and {Ballester}, O. and {Baltay}, C. and {Bault}, A. and {Beltran}, S.~F. and {Benavides}, B. and {BenZvi}, S. and {Berti}, A. and {Besuner}, R. and {Beutler}, Florian and {Bianchi}, D. and {Blake}, C. and {Blanc}, P. and {Blum}, R. and {Bolton}, A. and {Bose}, S. and {Bramall}, D. and {Brieden}, S. and {Brodzeller}, A. and {Brooks}, D. and {Brownewell}, C. and {Buckley-Geer}, E. and {Cahn}, R.~N. and {Cai}, Z. and {Canning}, R. and {Capasso}, R. and {Carnero Rosell}, A. and {Carton}, P. and {Casas}, R. and {Castander}, F.~J. and {Cervantes-Cota}, J.~L. and {Chabanier}, S. and {Chaussidon}, E. and {Chuang}, C. and {Circosta}, C. and {Cole}, S. and {Cooper}, A.~P. and {da Costa}, L. and {Cousinou}, M. -C. and {Cuceu}, A. and {Davis}, T.~M. and {Dawson}, K. and {de la Cruz-Noriega}, R. and {de la Macorra}, A. and {de Mattia}, A. and {Della Costa}, J. and {Demmer}, P. and {Derwent}, M. and {Dey}, A. and {Dey}, B. and {Dhungana}, G. and {Ding}, Z. and {Dobson}, C. and {Doel}, P. and {Donald-McCann}, J. and {Donaldson}, J. and {Douglass}, K. and {Duan}, Y. and {Dunlop}, P. and {Edelstein}, J. and {Eftekharzadeh}, S. and {Eisenstein}, D.~J. and {Enriquez-Vargas}, M. and {Escoffier}, S. and {Evatt}, M. and {Fagrelius}, P. and {Fan}, X. and {Fanning}, K. and {Fawcett}, V.~A. and {Ferraro}, S. and {Ereza}, J. and {Flaugher}, B. and {Font-Ribera}, A. and {Forero-Romero}, J.~E. and {Frenk}, C.~S. and {Fromenteau}, S. and {G{\"a}nsicke}, B.~T. and {Garcia-Quintero}, C. and {Garrison}, L. and {Gazta{\~n}aga}, E. and {Gerardi}, F. and {Gil-Mar{\'\i}n}, H. and {Gontcho A Gontcho}, S. and {Gonzalez-Morales}, Alma X. and {Gonzalez-de-Rivera}, G. and {Gonzalez-Perez}, V. and {Gordon}, C. and {Graur}, O. and {Green}, D. and {Grove}, C. and {Gruen}, D. and {Gutierrez}, G. and {Guy}, J. and {Hahn}, C. and {Harris}, S. and {Herrera}, D. and {Herrera-Alcantar}, Hiram K. and {Honscheid}, K. and {Howlett}, C. and {Huterer}, D. and {Ir{\v{s}}i{\v{c}}}, V. and {Ishak}, M. and {Jelinsky}, P. and {Jiang}, L. and {Jimenez}, J. and {Jing}, Y.~P. and {Joyce}, R. and {Jullo}, E. and {Juneau}, S. and {Kara{\c{c}}ayl{\i}}, N.~G. and {Karamanis}, M. and {Karcher}, A. and {Karim}, T. and {Kehoe}, R. and {Kent}, S. and {Kirkby}, D. and {Kisner}, T. and {Kitaura}, F. and {Koposov}, S.~E. and {Kov{\'a}cs}, A. and {Kremin}, A. and {Krolewski}, Alex and {L'Huillier}, B. and {Lahav}, O. and {Lambert}, A. and {Lamman}, C. and {Lan}, Ting-Wen and {Landriau}, M. and {Lane}, S. and {Lang}, D. and {Lange}, J.~U. and {Lasker}, J. and {Le Guillou}, L. and {Leauthaud}, A. and {Le Van Suu}, A. and {Levi}, Michael E. and {Li}, T.~S. and {Magneville}, C. and {Manera}, M. and {Manser}, Christopher J. and {Marshall}, B. and {Martini}, Paul and {McCollam}, W. and {McDonald}, P. and {Meisner}, Aaron M. and {Mena-Fern{\'a}ndez}, J. and {Meneses-Rizo}, J. and {Mezcua}, M. and {Miller}, T. and {Miquel}, R. and {Montero-Camacho}, P. and {Moon}, J. and {Moustakas}, J. and {Mueller}, E. and {Mu{\~n}oz-Guti{\'e}rrez}, Andrea and {Myers}, Adam D. and {Nadathur}, S. and {Najita}, J. and {Napolitano}, L. and {Neilsen}, E. and {Newman}, Jeffrey A. and {Nie}, J.~D. and {Ning}, Y. and {Niz}, G. and {Norberg}, P. and {Noriega}, Hern{\'a}n E. and {O'Brien}, T. and {Obuljen}, A. and {Palanque-Delabrouille}, N. and {Palmese}, A. and {Zhiwei}, P. and {Pappalardo}, D. and {PENG}, X. and {Percival}, W.~J. and {Perruchot}, S. and {Pogge}, R. and {Poppett}, C. and {Porredon}, A. and {Prada}, F. and {Prochaska}, J. and {Pucha}, R. and {P{\'e}rez-Fern{\'a}ndez}, A. and {P{\'e}rez-R{\`a}fols}, I. and {Rabinowitz}, D. and {Raichoor}, A. and {Ramirez-Solano}, S. and {Ram{\'\i}rez-P{\'e}rez}, C{\'e}sar and {Ravoux}, C. and {Reil}, K. and {Rezaie}, M. and {Rocher}, A. and {Rockosi}, C. and {Roe}, N.~A. and {Roodman}, A. and {Ross}, A.~J. and {Rossi}, G. and {Ruggeri}, R. and {Ruhlmann-Kleider}, V. and {Sabiu}, C.~G. and {Gaines}, S. and {Said}, K. and {Saintonge}, A. and {Salas Catonga}, Javier and {Samushia}, L. and {Sanchez}, E. and {Saulder}, C. and {Schaan}, E. and {Schlafly}, E. and {Schlegel}, D. and {Schmoll}, J. and {Scholte}, D. and {Schubnell}, M. and {Secroun}, A. and {Seo}, H. and {Serrano}, S. and {Sharples}, Ray M. and {Sholl}, Michael J. and {Silber}, Joseph Harry and {Silva}, D.~R. and {Sirk}, M. and {Siudek}, M. and {Smith}, A. and {Sprayberry}, D. and {Staten}, R. and {Stupak}, B. and {Tan}, T. and {Tarl{\'e}}, Gregory and {Tie}, Suk Sien and {Tojeiro}, R. and {Ure{\~n}a-L{\'o}pez}, L.~A. and {Valdes}, F. and {Valenzuela}, O. and {Valluri}, M. and {Vargas-Maga{\~n}a}, M. and {Verde}, L. and {Walther}, M. and {Wang}, B. and {Wang}, M.~S. and {Weaver}, B.~A. and {Weaverdyck}, C. and {Wechsler}, R. and {Wilson}, Michael J. and {Yang}, J. and {Yu}, Y. and {Yuan}, S. and {Y{\`e}che}, Christophe and {Zhang}, H. and {Zhang}, K. and {Zhao}, Cheng and {Zhou}, Rongpu and {Zhou}, Zhimin and {Zou}, H. and {Zou}, J. and {Zou}, S. and {Zu}, Y.},
        title = "{Overview of the Instrumentation for the Dark Energy Spectroscopic Instrument}",
      journal = {\aj},
         year = 2022,
        month = nov,
       volume = {164},
       number = {5},
          eid = {207},
        pages = {207},
          doi = {10.3847/1538-3881/ac882b},
archivePrefix = {arXiv},
       eprint = {2205.10939},
 primaryClass = {astro-ph.IM},
       adsurl = {https://ui.adsabs.harvard.edu/abs/2022AJ....164..207D}
}

@ARTICLE{2024Miller,
       author = {{Miller}, Timothy N. and {Doel}, Peter and {Gutierrez}, Gaston and {Besuner}, Robert and {Brooks}, David and {Gallo}, Giuseppe and {Heetderks}, Henry and {Jelinsky}, Patrick and {Kent}, Stephen M. and {Lampton}, Michael and {Levi}, Michael E. and {Liang}, Ming and {Meisner}, Aaron and {Sholl}, Michael J. and {Silber}, Joseph Harry and {Sprayberry}, David and {Aguilar}, Jessica Nicole and {de la Macorra}, Axel and {Eisenstein}, Daniel and {Fanning}, Kevin and {Font-Ribera}, Andreu and {Gazta{\~n}aga}, Enrique and {Gontcho A Gontcho}, Satya and {Honscheid}, Klaus and {Jimenez}, Jorge and {Joyce}, Dick and {Kehoe}, Robert and {Kisner}, Theodore and {Kremin}, Anthony and {Landriau}, Martin and {Le Guillou}, Laurent and {Magneville}, Christophe and {Martini}, Paul and {Miquel}, Ramon and {Moustakas}, John and {Nie}, Jundan and {Percival}, Will and {Poppett}, Claire and {Prada}, Francisco and {Rossi}, Graziano and {Schlegel}, David and {Schubnell}, Michael and {Seo}, Hee-Jong and {Sharples}, Ray and {Tarl{\'e}}, Gregory and {Vargas-Maga{\~n}a}, Mariana and {Zhou}, Zhimin and {the DESI Collaboration}},
        title = "{The Optical Corrector for the Dark Energy Spectroscopic Instrument}",
      journal = {\aj},
         year = 2024,
        month = aug,
       volume = {168},
       number = {2},
          eid = {95},
        pages = {95},
          doi = {10.3847/1538-3881/ad45fe},
archivePrefix = {arXiv},
       eprint = {2306.06310},
 primaryClass = {astro-ph.IM},
       adsurl = {https://ui.adsabs.harvard.edu/abs/2024AJ....168...95M}
}

@ARTICLE{2023Silber,
       author = {{Silber}, Joseph Harry and {Fagrelius}, Parker and {Fanning}, Kevin and {Schubnell}, Michael and {Aguilar}, Jessica Nicole and {Ahlen}, Steven and {Ameel}, Jon and {Ballester}, Otger and {Baltay}, Charles and {Bebek}, Chris and {Benton Beard}, Dominic and {Besuner}, Robert and {Cardiel-Sas}, Laia and {Casas}, Ricard and {Castander}, Francisco Javier and {Claybaugh}, Todd and {Dobson}, Carl and {Duan}, Yutong and {Dunlop}, Patrick and {Edelstein}, Jerry and {Emmet}, William T. and {Elliott}, Ann and {Evatt}, Matthew and {Gershkovich}, Irena and {Guy}, Julien and {Harris}, Stu and {Heetderks}, Henry and {Heetderks}, Ian and {Honscheid}, Klaus and {Illa}, Jose Maria and {Jelinsky}, Patrick and {Jelinsky}, Sharon R. and {Jimenez}, Jorge and {Karcher}, Armin and {Kent}, Stephen and {Kirkby}, David and {Kneib}, Jean-Paul and {Lambert}, Andrew and {Lampton}, Mike and {Leitner}, Daniela and {Levi}, Michael and {McCauley}, Jeremy and {Meisner}, Aaron and {Miller}, Timothy N. and {Miquel}, Ramon and {Mundet}, Juli{\'a} and {Poppett}, Claire and {Rabinowitz}, David and {Reil}, Kevin and {Roman}, David and {Schlegel}, David and {Serrano}, Santiago and {Van Shourt}, William and {Sprayberry}, David and {Tarl{\'e}}, Gregory and {Tie}, Suk Sien and {Weaverdyck}, Curtis and {Zhang}, Kai and {Azzaro}, Marco and {Bailey}, Stephen and {Becerril}, Santiago and {Blackwell}, Tami and {Bouri}, Mohamed and {Brooks}, David and {Buckley-Geer}, Elizabeth and {Castro}, Jose Pe{\~n}ate and {Derwent}, Mark and {Dey}, Arjun and {Dhungana}, Govinda and {Doel}, Peter and {Eisenstein}, Daniel J. and {Fahim}, Nasib and {Garcia-Bellido}, Juan and {Gazta{\~n}aga}, Enrique and {A Gontcho}, Satya Gontcho and {Gutierrez}, Gaston and {H{\"o}rler}, Philipp and {Kehoe}, Robert and {Kisner}, Theodore and {Kremin}, Anthony and {Kronig}, Luzius and {Landriau}, Martin and {Le Guillou}, Laurent and {Martini}, Paul and {Moustakas}, John and {Palanque-Delabrouille}, Nathalie and {Peng}, Xiyan and {Percival}, Will and {Prada}, Francisco and {Allende Prieto}, Carlos and {de Rivera}, Guillermo Gonzalez and {Sanchez}, Eusebio and {Sanchez}, Justo and {Sharples}, Ray and {Soares-Santos}, Marcelle and {Schlafly}, Edward and {Weaver}, Benjamin Alan and {Zhou}, Zhimin and {Zhu}, Yaling and {Zou}, Hu and {DESI Collaboration}},
        title = "{The Robotic Multiobject Focal Plane System of the Dark Energy Spectroscopic Instrument (DESI)}",
      journal = {\aj},
         year = 2023,
        month = jan,
       volume = {165},
       number = {1},
          eid = {9},
        pages = {9},
          doi = {10.3847/1538-3881/ac9ab1},
archivePrefix = {arXiv},
       eprint = {2205.09014},
 primaryClass = {astro-ph.IM},
       adsurl = {https://ui.adsabs.harvard.edu/abs/2023AJ....165....9S}
}

@ARTICLE{Zhou2020,
       author = {{Zhou}, Rongpu and {Newman}, Jeffrey A. and {Dawson}, Kyle S. and {Eisenstein}, Daniel J. and {Brooks}, David D. and {Dey}, Arjun and {Dey}, Biprateep and {Duan}, Yutong and {Eftekharzadeh}, Sarah and {Gazta{\~n}aga}, Enrique and {Kehoe}, Robert and {Landriau}, Martin and {Levi}, Michael E. and {Licquia}, Timothy C. and {Meisner}, Aaron M. and {Moustakas}, John and {Myers}, Adam D. and {Palanque-Delabrouille}, Nathalie and {Poppett}, Claire and {Prada}, Francisco and {Raichoor}, Anand and {Schlegel}, David J. and {Schubnell}, Michael and {Staten}, Ryan and {Tarl{\'e}}, Gregory and {Y{\`e}che}, Christophe},
        title = "{Preliminary Target Selection for the DESI Luminous Red Galaxy (LRG) Sample}",
      journal = {Research Notes of the American Astronomical Society},
         year = 2020,
        month = oct,
       volume = {4},
       number = {10},
          eid = {181},
        pages = {181},
          doi = {10.3847/2515-5172/abc0f4},
archivePrefix = {arXiv},
       eprint = {2010.11282},
 primaryClass = {astro-ph.CO},
       adsurl = {https://ui.adsabs.harvard.edu/abs/2020RNAAS...4..181Z}
}

@ARTICLE{Zhou2023,
       author = {{Zhou}, Rongpu and {Dey}, Biprateep and {Newman}, Jeffrey A. and {Eisenstein}, Daniel J. and {Dawson}, K. and {Bailey}, S. and {Berti}, A. and {Guy}, J. and {Lan}, Ting-Wen and {Zou}, H. and {Aguilar}, J. and {Ahlen}, S. and {Alam}, Shadab and {Brooks}, D. and {de la Macorra}, A. and {Dey}, A. and {Dhungana}, G. and {Fanning}, K. and {Font-Ribera}, A. and {Gontcho}, S. Gontcho A. and {Honscheid}, K. and {Ishak}, Mustapha and {Kisner}, T. and {Kov{\'a}cs}, A. and {Kremin}, A. and {Landriau}, M. and {Levi}, Michael E. and {Magneville}, C. and {Manera}, Marc and {Martini}, P. and {Meisner}, Aaron M. and {Miquel}, R. and {Moustakas}, J. and {Myers}, Adam D. and {Nie}, Jundan and {Palanque-Delabrouille}, N. and {Percival}, W.~J. and {Poppett}, C. and {Prada}, F. and {Raichoor}, A. and {Ross}, A.~J. and {Schlafly}, E. and {Schlegel}, D. and {Schubnell}, M. and {Tarl{\'e}}, Gregory and {Weaver}, B.~A. and {Wechsler}, R.~H. and {Y{\'e}che}, Christophe and {Zhou}, Zhimin},
        title = "{Target Selection and Validation of DESI Luminous Red Galaxies}",
      journal = {\aj},
         year = 2023,
        month = feb,
       volume = {165},
       number = {2},
          eid = {58},
        pages = {58},
          doi = {10.3847/1538-3881/aca5fb},
archivePrefix = {arXiv},
       eprint = {2208.08515},
 primaryClass = {astro-ph.CO},
       adsurl = {https://ui.adsabs.harvard.edu/abs/2023AJ....165...58Z}
}

@ARTICLE{Raichoor2020,
       author = {{Raichoor}, Anand and {Eisenstein}, Daniel J. and {Karim}, Tanveer and {Newman}, Jeffrey A. and {Moustakas}, John and {Brooks}, David D. and {Dawson}, Kyle S. and {Dey}, Arjun and {Duan}, Yutong and {Eftekharzadeh}, Sarah and {Gazta{\~n}aga}, Enrique and {Kehoe}, Robert and {Landriau}, Martin and {Lang}, Dustin and {Lee}, Jae H. and {Levi}, Michael E. and {Meisner}, Aaron M. and {Myers}, Adam D. and {Palanque-Delabrouille}, Nathalie and {Poppett}, Claire and {Prada}, Francisco and {Ross}, Ashley J. and {Schlegel}, David J. and {Schubnell}, Michael and {Staten}, Ryan and {Tarl{\'e}}, Gregory and {Tojeiro}, Rita and {Y{\`e}che}, Christophe and {Zhou}, Rongpu},
        title = "{Preliminary Target Selection for the DESI Emission Line Galaxy (ELG) Sample}",
      journal = {Research Notes of the American Astronomical Society},
         year = 2020,
        month = oct,
       volume = {4},
       number = {10},
          eid = {180},
        pages = {180},
          doi = {10.3847/2515-5172/abc078},
archivePrefix = {arXiv},
       eprint = {2010.11281},
 primaryClass = {astro-ph.CO},
       adsurl = {https://ui.adsabs.harvard.edu/abs/2020RNAAS...4..180R}
}

@ARTICLE{Raichoor2023AJ,
       author = {{Raichoor}, A. and {Moustakas}, J. and {Newman}, Jeffrey A. and {Karim}, T. and {Ahlen}, S. and {Alam}, Shadab and {Bailey}, S. and {Brooks}, D. and {Dawson}, K. and {de la Macorra}, A. and {de Mattia}, A. and {Dey}, A. and {Dey}, Biprateep and {Dhungana}, G. and {Eftekharzadeh}, S. and {Eisenstein}, D.~J. and {Fanning}, K. and {Font-Ribera}, A. and {Garc{\'\i}a-Bellido}, J. and {Gazta{\~n}aga}, E. and {A Gontcho}, S. Gontcho and {Guy}, J. and {Honscheid}, K. and {Ishak}, M. and {Kehoe}, R. and {Kisner}, T. and {Kremin}, Anthony and {Lan}, Ting-Wen and {Landriau}, M. and {Le Guillou}, L. and {Levi}, Michael E. and {Magneville}, C. and {Manera}, M. and {Martini}, P. and {Meisner}, Aaron M. and {Myers}, Adam D. and {Nie}, Jundan and {Palanque-Delabrouille}, N. and {Percival}, W.~J. and {Poppett}, C. and {Prada}, F. and {Ross}, A.~J. and {Ruhlmann-Kleider}, V. and {Sabiu}, C.~G. and {Schlafly}, E.~F. and {Schlegel}, D. and {Tarl{\'e}}, Gregory and {Weaver}, B.~A. and {Y{\`e}che}, Christophe and {Zhou}, Rongpu and {Zhou}, Zhimin and {Zou}, H.},
        title = "{Target Selection and Validation of DESI Emission Line Galaxies}",
      journal = {\aj},
         year = 2023,
        month = mar,
       volume = {165},
       number = {3},
          eid = {126},
        pages = {126},
          doi = {10.3847/1538-3881/acb213},
archivePrefix = {arXiv},
       eprint = {2208.08513},
 primaryClass = {astro-ph.CO},
       adsurl = {https://ui.adsabs.harvard.edu/abs/2023AJ....165..126R}
}

@ARTICLE{Yeche2020,
       author = {{Y{\`e}che}, Christophe and {Palanque-Delabrouille}, Nathalie and {Claveau}, Charles-Antoine and {Brooks}, David D. and {Chaussidon}, Edmond and {Davis}, Tamara M. and {Dawson}, Kyle S. and {Dey}, Arjun and {Duan}, Yutong and {Eftekharzadeh}, Sarah and {Eisenstein}, Daniel J. and {Gazta{\~n}aga}, Enrique and {Kehoe}, Robert and {Landriau}, Martin and {Lang}, Dustin and {Levi}, Michael E. and {Meisner}, Aaron M. and {Myers}, Adam D. and {Newman}, Jeffrey A. and {Poppett}, Claire and {Prada}, Francisco and {Raichoor}, Anand and {Schlegel}, David J. and {Schubnell}, Michael and {Staten}, Ryan and {Tarl{\'e}}, Gregory and {Zhou}, Rongpu},
        title = "{Preliminary Target Selection for the DESI Quasar (QSO) Sample}",
      journal = {Research Notes of the American Astronomical Society},
         year = 2020,
        month = oct,
       volume = {4},
       number = {10},
          eid = {179},
        pages = {179},
          doi = {10.3847/2515-5172/abc01a},
archivePrefix = {arXiv},
       eprint = {2010.11280},
 primaryClass = {astro-ph.CO},
       adsurl = {https://ui.adsabs.harvard.edu/abs/2020RNAAS...4..179Y}
}

@ARTICLE{Chaussidon2023,
       author = {{Chaussidon}, Edmond and {Y{\`e}che}, Christophe and {Palanque-Delabrouille}, Nathalie and {Alexander}, David M. and {Yang}, Jinyi and {Ahlen}, Steven and {Bailey}, Stephen and {Brooks}, David and {Cai}, Zheng and {Chabanier}, Sol{\`e}ne and {Davis}, Tamara M. and {Dawson}, Kyle and {de laMacorra}, Axel and {Dey}, Arjun and {Dey}, Biprateep and {Eftekharzadeh}, Sarah and {Eisenstein}, Daniel J. and {Fanning}, Kevin and {Font-Ribera}, Andreu and {Gazta{\~n}aga}, Enrique and {A Gontcho}, Satya Gontcho and {Gonzalez-Morales}, Alma X. and {Guy}, Julien and {Herrera-Alcantar}, Hiram K. and {Honscheid}, Klaus and {Ishak}, Mustapha and {Jiang}, Linhua and {Juneau}, Stephanie and {Kehoe}, Robert and {Kisner}, Theodore and {Kov{\'a}cs}, Andras and {Kremin}, Anthony and {Lan}, Ting-Wen and {Landriau}, Martin and {Le Guillou}, Laurent and {Levi}, Michael E. and {Magneville}, Christophe and {Martini}, Paul and {Meisner}, Aaron M. and {Moustakas}, John and {Mu{\~n}oz-Guti{\'e}rrez}, Andrea and {Myers}, Adam D. and {Newman}, Jeffrey A. and {Nie}, Jundan and {Percival}, Will J. and {Poppett}, Claire and {Prada}, Francisco and {Raichoor}, Anand and {Ravoux}, Corentin and {Ross}, Ashley J. and {Schlafly}, Edward and {Schlegel}, David and {Tan}, Ting and {Tarl{\'e}}, Gregory and {Zhou}, Rongpu and {Zhou}, Zhimin and {Zou}, Hu},
        title = "{Target Selection and Validation of DESI Quasars}",
      journal = {\apj},
         year = 2023,
        month = feb,
       volume = {944},
       number = {1},
          eid = {107},
        pages = {107},
          doi = {10.3847/1538-4357/acb3c2},
archivePrefix = {arXiv},
       eprint = {2208.08511},
 primaryClass = {astro-ph.CO},
       adsurl = {https://ui.adsabs.harvard.edu/abs/2023ApJ...944..107C}
}

@ARTICLE{Guy2023,
       author = {{Guy}, J. and {Bailey}, S. and {Kremin}, A. and {Alam}, Shadab and {Alexander}, D.~M. and {Allende Prieto}, C. and {BenZvi}, S. and {Bolton}, A.~S. and {Brooks}, D. and {Chaussidon}, E. and {Cooper}, A.~P. and {Dawson}, K. and {de la Macorra}, A. and {Dey}, A. and {Dey}, Biprateep and {Dhungana}, G. and {Eisenstein}, D.~J. and {Font-Ribera}, A. and {Forero-Romero}, J.~E. and {Gazta{\~n}aga}, E. and {Gontcho A Gontcho}, S. and {Green}, D. and {Honscheid}, K. and {Ishak}, M. and {Kehoe}, R. and {Kirkby}, D. and {Kisner}, T. and {Koposov}, Sergey E. and {Lan}, Ting-Wen and {Landriau}, M. and {Le Guillou}, L. and {Levi}, Michael E. and {Magneville}, C. and {Manser}, Christopher J. and {Martini}, P. and {Meisner}, Aaron M. and {Miquel}, R. and {Moustakas}, J. and {Myers}, Adam D. and {Newman}, Jeffrey A. and {Nie}, Jundan and {Palanque-Delabrouille}, N. and {Percival}, W.~J. and {Poppett}, C. and {Prada}, F. and {Raichoor}, A. and {Ravoux}, C. and {Ross}, A.~J. and {Schlafly}, E.~F. and {Schlegel}, D. and {Schubnell}, M. and {Sharples}, Ray M. and {Tarl{\'e}}, Gregory and {Weaver}, B.~A. and {Y{\'e}che}, Christophe and {Zhou}, Rongpu and {Zhou}, Zhimin and {Zou}, H.},
        title = "{The Spectroscopic Data Processing Pipeline for the Dark Energy Spectroscopic Instrument}",
      journal = {\aj},
         year = 2023,
        month = apr,
       volume = {165},
       number = {4},
          eid = {144},
        pages = {144},
          doi = {10.3847/1538-3881/acb212},
archivePrefix = {arXiv},
       eprint = {2209.14482},
 primaryClass = {astro-ph.IM},
       adsurl = {https://ui.adsabs.harvard.edu/abs/2023AJ....165..144G}
}

@ARTICLE{Zou2024,
       author = {{Zou}, Hu and {Sui}, Jipeng and {Saintonge}, Am{\'e}lie and {Scholte}, Dirk and {Moustakas}, John and {Siudek}, Malgorzata and {Dey}, Arjun and {Juneau}, Stephanie and {Guo}, Weijian and {Canning}, Rebecca and {Aguilar}, J. and {Ahlen}, S. and {Brooks}, D. and {Claybaugh}, T. and {Dawson}, K. and {de la Macorra}, A. and {Doel}, P. and {Forero-Romero}, J.~E. and {Gontcho A Gontcho}, S. and {Honscheid}, K. and {Landriau}, M. and {Le Guillou}, L. and {Manera}, M. and {Meisner}, A. and {Miquel}, R. and {Nie}, Jundan and {Poppett}, C. and {Rezaie}, M. and {Rossi}, G. and {Sanchez}, E. and {Schubnell}, M. and {Seo}, H. and {Tarl{\'e}}, G. and {Zhou}, Zhimin and {Zou}, Siwei},
        title = "{A Large Sample of Extremely Metal-poor Galaxies at z < 1 Identified from the DESI Early Data}",
      journal = {\apj},
         year = 2024,
        month = feb,
       volume = {961},
       number = {2},
          eid = {173},
        pages = {173},
          doi = {10.3847/1538-4357/ad1409},
archivePrefix = {arXiv},
       eprint = {2312.00300},
 primaryClass = {astro-ph.GA},
       adsurl = {https://ui.adsabs.harvard.edu/abs/2024ApJ...961..173Z}
}

@ARTICLE{Brammer2008dispersion,
       author = {{Brammer}, Gabriel B. and {van Dokkum}, Pieter G. and {Coppi}, Paolo},
        title = "{EAZY: A Fast, Public Photometric Redshift Code}",
      journal = {\apj},
         year = 2008,
        month = oct,
       volume = {686},
       number = {2},
        pages = {1503-1513},
          doi = {10.1086/591786},
archivePrefix = {arXiv},
       eprint = {0807.1533},
 primaryClass = {astro-ph},
       adsurl = {https://ui.adsabs.harvard.edu/abs/2008ApJ...686.1503B}
}

@ARTICLE{Wojtak2011Nature,
       author = {{Wojtak}, Rados{\l}aw and {Hansen}, Steen H. and {Hjorth}, Jens},
        title = "{Gravitational redshift of galaxies in clusters as predicted by general relativity}",
      journal = {\nat},
         year = 2011,
        month = sep,
       volume = {477},
       number = {7366},
        pages = {567-569},
          doi = {10.1038/nature10445},
archivePrefix = {arXiv},
       eprint = {1109.6571},
 primaryClass = {astro-ph.CO},
       adsurl = {https://ui.adsabs.harvard.edu/abs/2011Natur.477..567W}
}

@ARTICLE{Strateva2001bimodal,
       author = {{Strateva}, Iskra and {Ivezi{\'c}}, {\v{Z}}eljko and {Knapp}, Gillian R. and {Narayanan}, Vijay K. and {Strauss}, Michael A. and {Gunn}, James E. and {Lupton}, Robert H. and {Schlegel}, David and {Bahcall}, Neta A. and {Brinkmann}, Jon and {Brunner}, Robert J. and {Budav{\'a}ri}, Tam{\'a}s and {Csabai}, Istv{\'a}n and {Castander}, Francisco Javier and {Doi}, Mamoru and {Fukugita}, Masataka and {Gy{\H{o}}ry}, Zsuzsanna and {Hamabe}, Masaru and {Hennessy}, Greg and {Ichikawa}, Takashi and {Kunszt}, Peter Z. and {Lamb}, Don Q. and {McKay}, Timothy A. and {Okamura}, Sadanori and {Racusin}, Judith and {Sekiguchi}, Maki and {Schneider}, Donald P. and {Shimasaku}, Kazuhiro and {York}, Donald},
        title = "{Color Separation of Galaxy Types in the Sloan Digital Sky Survey Imaging Data}",
      journal = {\aj},
         year = 2001,
        month = oct,
       volume = {122},
       number = {4},
        pages = {1861-1874},
          doi = {10.1086/323301},
archivePrefix = {arXiv},
       eprint = {astro-ph/0107201},
 primaryClass = {astro-ph},
       adsurl = {https://ui.adsabs.harvard.edu/abs/2001AJ....122.1861S}
}

@ARTICLE{Williams2009bimodal,
       author = {{Williams}, Rik J. and {Quadri}, Ryan F. and {Franx}, Marijn and {van Dokkum}, Pieter and {Labb{\'e}}, Ivo},
        title = "{Detection of Quiescent Galaxies in a Bicolor Sequence from Z = 0-2}",
      journal = {\apj},
         year = 2009,
        month = feb,
       volume = {691},
       number = {2},
        pages = {1879-1895},
          doi = {10.1088/0004-637X/691/2/1879},
archivePrefix = {arXiv},
       eprint = {0806.0625},
 primaryClass = {astro-ph},
       adsurl = {https://ui.adsabs.harvard.edu/abs/2009ApJ...691.1879W}
}

@ARTICLE{BC03,
       author = {{Bruzual}, G. and {Charlot}, S.},
        title = "{Stellar population synthesis at the resolution of 2003}",
      journal = {\mnras},
         year = 2003,
        month = oct,
       volume = {344},
       number = {4},
        pages = {1000-1028},
          doi = {10.1046/j.1365-8711.2003.06897.x},
archivePrefix = {arXiv},
       eprint = {astro-ph/0309134},
 primaryClass = {astro-ph},
       adsurl = {https://ui.adsabs.harvard.edu/abs/2003MNRAS.344.1000B}
}

@ARTICLE{commah,
       author = {{Correa}, Camila A. and {Wyithe}, J. Stuart B. and {Schaye}, Joop and {Duffy}, Alan R.},
        title = "{The accretion history of dark matter haloes - I. The physical origin of the universal function}",
      journal = {\mnras},
         year = 2015,
        month = jun,
       volume = {450},
       number = {2},
        pages = {1514-1520},
          doi = {10.1093/mnras/stv689},
archivePrefix = {arXiv},
       eprint = {1409.5228},
 primaryClass = {astro-ph.GA},
       adsurl = {https://ui.adsabs.harvard.edu/abs/2015MNRAS.450.1514C}
}

@ARTICLE{2015commah,
       author = {{Correa}, Camila A. and {Wyithe}, J. Stuart B. and {Schaye}, Joop and {Duffy}, Alan R.},
        title = "{The accretion history of dark matter haloes - II. The connections with the mass power spectrum and the density profile}",
      journal = {\mnras},
         year = 2015,
        month = jun,
       volume = {450},
       number = {2},
        pages = {1521-1537},
          doi = {10.1093/mnras/stv697},
archivePrefix = {arXiv},
       eprint = {1501.04382},
 primaryClass = {astro-ph.CO},
       adsurl = {https://ui.adsabs.harvard.edu/abs/2015MNRAS.450.1521C}
}

@ARTICLE{2015commah1,
       author = {{Correa}, Camila A. and {Wyithe}, J. Stuart B. and {Schaye}, Joop and {Duffy}, Alan R.},
        title = "{The accretion history of dark matter haloes - III. A physical model for the concentration-mass relation}",
      journal = {\mnras},
         year = 2015,
        month = sep,
       volume = {452},
       number = {2},
        pages = {1217-1232},
          doi = {10.1093/mnras/stv1363},
archivePrefix = {arXiv},
       eprint = {1502.00391},
 primaryClass = {astro-ph.CO},
       adsurl = {https://ui.adsabs.harvard.edu/abs/2015MNRAS.452.1217C}
}

@ARTICLE{2011BCG,
       author = {{Pipino}, A. and {Szabo}, T. and {Pierpaoli}, E. and {MacKenzie}, S.~M. and {Dong}, F.},
        title = "{The properties of brightest cluster galaxies in the Sloan Digital Sky Survey Data Release 6 adaptive matched filter cluster catalogue}",
      journal = {\mnras},
         year = 2011,
        month = nov,
       volume = {417},
       number = {4},
        pages = {2817-2830},
          doi = {10.1111/j.1365-2966.2011.19444.x},
archivePrefix = {arXiv},
       eprint = {1011.3017},
 primaryClass = {astro-ph.CO},
       adsurl = {https://ui.adsabs.harvard.edu/abs/2011MNRAS.417.2817P}
}

@ARTICLE{2019BCG,
       author = {{Cerulo}, P. and {Orellana}, G.~A. and {Covone}, G.},
        title = "{The evolution of brightest cluster galaxies in the nearby Universe - I. Colours and stellar masses from the Sloan Digital Sky Survey and Wide Infrared Survey Explorer}",
      journal = {\mnras},
         year = 2019,
        month = aug,
       volume = {487},
       number = {3},
        pages = {3759-3775},
          doi = {10.1093/mnras/stz1495},
archivePrefix = {arXiv},
       eprint = {1905.12117},
 primaryClass = {astro-ph.GA},
       adsurl = {https://ui.adsabs.harvard.edu/abs/2019MNRAS.487.3759C}
}

@ARTICLE{2024chen,
       author = {{Chen}, Zhaobin and {Gu}, Yizhou and {Zou}, Hu and {Yuan}, Qirong},
        title = "{Galaxy Clusters from the DESI Legacy Imaging Surveys. II. Environmental Effects on the Size{\textendash}Mass Relation}",
      journal = {\apj},
         year = 2024,
        month = feb,
       volume = {961},
       number = {2},
          eid = {253},
        pages = {253},
          doi = {10.3847/1538-4357/ad15fd},
archivePrefix = {arXiv},
       eprint = {2312.17075},
 primaryClass = {astro-ph.GA},
       adsurl = {https://ui.adsabs.harvard.edu/abs/2024ApJ...961..253C}
}

@ARTICLE{2022BCG_important,
       author = {{Orellana-Gonz{\'a}lez}, G. and {Cerulo}, P. and {Covone}, G. and {Cheng}, C. and {Leiton}, R. and {Demarco}, R. and {Gendron-Marsolais}, M. -L.},
        title = "{The evolution of brightest cluster galaxies in the nearby Universe II: The star-formation activity and the stellar mass from spectral energy distribution}",
      journal = {\mnras},
         year = 2022,
        month = may,
       volume = {512},
       number = {2},
        pages = {2758-2776},
          doi = {10.1093/mnras/stac001},
archivePrefix = {arXiv},
       eprint = {2201.00826},
 primaryClass = {astro-ph.GA},
       adsurl = {https://ui.adsabs.harvard.edu/abs/2022MNRAS.512.2758O}
}

@ARTICLE{2012bimodality,
       author = {{Wetzel}, Andrew R. and {Tinker}, Jeremy L. and {Conroy}, Charlie},
        title = "{Galaxy evolution in groups and clusters: star formation rates, red sequence fractions and the persistent bimodality}",
      journal = {\mnras},
         year = 2012,
        month = jul,
       volume = {424},
       number = {1},
        pages = {232-243},
          doi = {10.1111/j.1365-2966.2012.21188.x},
archivePrefix = {arXiv},
       eprint = {1107.5311},
 primaryClass = {astro-ph.CO},
       adsurl = {https://ui.adsabs.harvard.edu/abs/2012MNRAS.424..232W}
}





\appendix

\section{Correlation Between Cluster Halo Mass and BCG Stellar Mass}\label{appendix A}

\begin{figure}
    \includegraphics[width=\columnwidth]{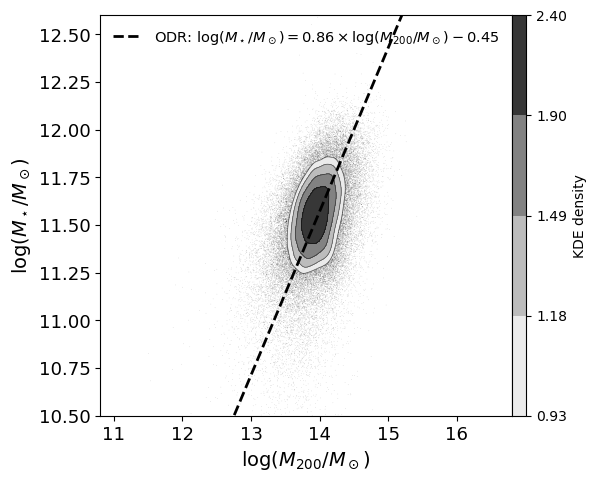}
    \centering
    \caption{The correlation between BCG stellar mass and cluster halo mass in our sample.}
    \label{fig:halo-mass-all}
\end{figure}

Previous studies \citep{2015halo,2016halo, 2019halo} have established a positive correlation between BCG stellar mass and cluster halo mass. Using our sample, we confirm this trend: Figure~\ref{fig:halo-mass-all} shows a clear relation between BCG stellar mass and halo mass, with more massive haloes hosting more massive BCGs. In our analysis, we apply Orthogonal Distance Regression (ODR) to quantify the correlation between BCG stellar mass ($\log(M_{\star}/M_{\odot})$) and halo mass ($\log(M_{\mathrm{200}}/M_{\odot})$), which yields a slope of 0.86. This ODR slope is steeper than the power-law index reported by \cite{2016halo} with $\alpha = 0.64$, potentially reflecting differences in sample selection. Relative to \cite{2019halo}, who found slopes of 0.41 at low redshift ($0.1 \leq z \leq 0.3$) and 0.31 at higher redshift ($0.3 < z \leq 0.65$), our ODR result is steeper than both bins. Overall, the ODR analysis reinforces the view that more massive haloes host more massive BCGs, while highlighting modest sample-dependent variations in the inferred scaling.

\citep{test}

\bsp	
\label{lastpage}
\end{document}